\documentclass[prl,reprint,twocolumn,superscriptaddress,floatfix,nofootinbib,longbibliography]{revtex4-1}
\usepackage{epsfig,amsmath,amssymb,color,comment,physics}
\usepackage[makeroom]{cancel}
\usepackage[caption=false]{subfig}
\usepackage{mathrsfs}
\usepackage[countmax]{subfloat}
\usepackage[normalem]{ulem}
\usepackage[english]{babel}
\usepackage{dsfont}
\usepackage{braket}
\usepackage[bookmarks=true,colorlinks,linkcolor=OrangeRed,urlcolor=NavyBlue,citecolor=RoyalBlue]{hyperref}
\usepackage[dvipsnames]{xcolor}
\usepackage{graphicx}
\usepackage{cleveref}
\graphicspath{{./Figures/}}
\usepackage{amsmath, amssymb,latexsym,amsfonts}

\newcommand{\h}[1]{\hat#1}
\pdfoutput=1
\begin{document}
\title{Weak universality induced by $Q=\pm 2e$ charges at the deconfinement
transition of a (2+1)-d $U(1)$ lattice gauge theory} 

\author{Indrajit Sau}
\affiliation{School of Physical Sciences, Indian Association for the Cultivation 
of Science, Kolkata 700032,India}

\author{Arnab Sen}
\affiliation{School of Physical Sciences, Indian Association for the Cultivation 
of Science, Kolkata 700032,India}

\author{Debasish Banerjee}
\affiliation{Theory Division, Saha Institute of Nuclear Physics, 1/AF Bidhan Nagar, 
Kolkata 700064, India}
\affiliation{Homi Bhabha National Institute, Training School Complex, Anushaktinagar, 
Mumbai 400094,India}

\begin{abstract}
  Matter-free lattice gauge theories (LGTs) provide an ideal setting to understand 
  confinement to deconfinement transitions at finite temperatures, which is typically
  due to the spontaneous breakdown (at large temperatures) of the centre symmetry 
  associated with the gauge group. Close to the transition, the relevant degrees of 
  freedom (Polyakov loop) transform under these centre symmetries, and the effective 
  theory only depends on the Polyakov loop and its fluctuations. As shown first 
  by Svetitsky and Yaffe, and subsequently verified numerically, for the $U(1)$ LGT 
  in $(2+1)$-d the transition is in the 2-d XY universality class, while for the 
  $Z_2$ LGT, it is in the 2-d Ising universality class. We extend this classic scenario 
  by adding higher charged matter fields, and show that the notion of universality 
  is generalized such that the critical exponents $\gamma, \nu$ can change continuously 
  as a coupling is varied, while their ratio is fixed to the 2-d Ising value. While 
  such weak universality is well-known for spin models, we demonstrate this for LGTs 
  for the first time. Using an efficient cluster algorithm, we show that the finite 
  temperature phase transition of the $U(1)$ quantum link LGT in the spin $S=\frac{1}{2}$ 
  representation is in the 2-d XY universality class, as expected. On the addition 
  of $Q = \pm 2e$ charges distributed thermally, we demonstrate the occurrence of 
  weak universality.   
\end{abstract}

\date{\today}
\maketitle

\paragraph{Introduction.--} Phases of matter at extreme physical conditions of temperature, 
pressure, and density often challenge our conventional notions and stimulate extensive 
research, both experimentally and theoretically. Of particular relevance are the physics 
of the early universe, and the interior of neutron stars. Both scenarios are expected to 
have a microscopic description through quantum chromodynamics (QCD), a field theory of 
quarks and gluons \cite{Wilczek1999,Philipsen2019,DElia2018,Sharma2021}. As a quantum 
field theory (QFT), QCD is a strongly interacting theory which confines colour-charge 
carrying quarks and gluons into colour singlet bound states, through the confinement 
phenomenon rendering conventional perturbation techniques unsuitable. 

 Lattice gauge theories (LGT) are non-perturbative formulation of QFTs,  where Markov chain 
Monte Carlo methods are used to compute expectation values of physical observables, and 
supply the most reliable insights about the strong interaction physics \cite{USQCD2022}. 
The possibility of a phase transition out of the confined phase at finite temperatures was 
explored first using the computationally simpler case of matter-free pure gauge theories. 
It is universally accepted that pure $SU(3)$ gauge theory has a first order deconfinement 
phase transition \cite{Creutz1982, Kogut1984},  while the $SU(2)$ gauge theory has a second 
order phase transition \cite{Creutz1980, Gavai1982}. For QCD with physical quark masses, 
there is only a crossover from the hadronic phase to the deconfined quark gluon plasma phase
\cite{Hasenfratz1983, Bhattacharya2014}. 

\begin{figure}
    \hspace{-0.5cm}
    \includegraphics[width=4cm, height=4cm]{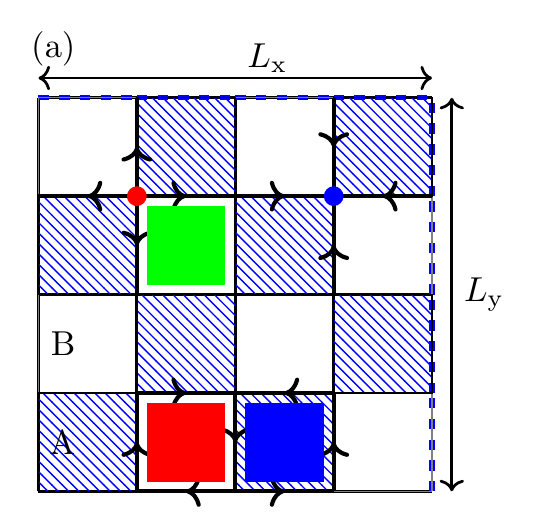}
    \hspace{-0.5cm}
    \includegraphics[width=5cm, height=4cm]{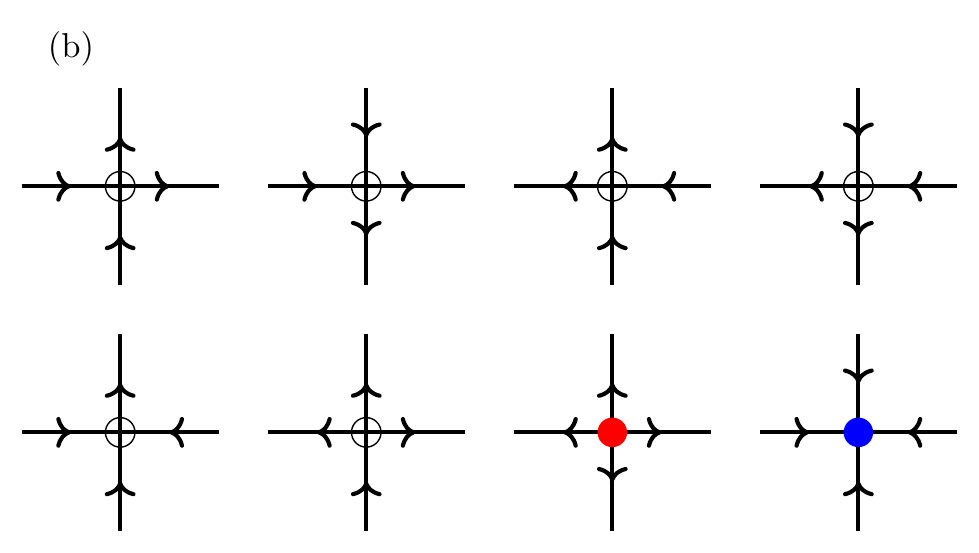}
    \caption{(a) Sketch of the lattice, the dashed and clear plaquettes denote the A and B sublattice
    respectively. $Q = \pm 2e$ charges are shown as red and blue circles, respectively.
    Plaquettes can be non-flippable (shaded in green), especially when it has a charge $\pm 2e$ at a 
    corner, or flippable in the clockwise (red) or anti-clockwise (blue) sense. (b) The different Gauss
    law realizations which contribute with equal weight, the first six have $Q=0$, while the last 
    two have $Q = \pm 2e$ at the vertex.}
    \label{fig:setup}
\end{figure}

  The finite temperature phase transition in pure gauge theories can be described using an effective 
field theory (EFT). Svetitsky and Yaffe (SY) \cite{Svetitsky1982} used the insight that the 
confinement to deconfinement transition in a pure gauge theory is due to the spontaneous breakdown of 
the global centre symmetry of the gauge group to show the relevant degrees of freedom in the EFT 
are the Polyakov loop (order parameter), and its fluctuations. Integrating out all other degrees 
of freedom in the original gauge theory in $(d+1)$-dimensions, they argued that the EFT corresponds 
to a $d$-dimensional scalar field theory or spin system. The confinement in the original gauge theory 
ensures that the effective couplings in the spin system are all short ranged. Using universality, 
they argued that if the effective spin model has a second order transition, then the original gauge 
theory should also have it. This SY scenario has been verified in different numerical simulations
\cite{Caselle1995,Bonati2013,Lau2015,Borisenko2015,Biswal2016,Kuramashi2018}, and is widely regarded 
as a success of universality. 

  In this Letter, we report an extension of the SY approach of understanding the finite temperature
phase transition of a (2+1)-d $U(1)$ lattice gauge theory in the presence of higher charges $\pm 2e$ 
(where $e$ denotes the fundamental unit of charge, which we set to $1$ henceforth). As we will 
demonstrate using a specific example of a quantum link model (QLM), the presence of the even charges 
keeps intact only a global $Z_2$ subgroup of the full $U(1)$ centre symmetry. Consequently, the SY 
prediction of a Berezinskii-Kosterlitz-Thouless (BKT) for the deconfinement transition of the pure 
$U(1)$ LGT gets modified. Instead, a weak universality phenomenon arises \cite{Suzuki1974}, characterized 
by large values of critical exponents like $\gamma, \nu, \beta$, while others like $\eta, \delta$ defined 
directly at the critical point, 
as well as the ratios of the critical exponents are fixed to the 2-d $Z_2$ Ising model. Weak universality 
has been observed in a variety of systems 
\cite{Baxter1971, Ashkin1943,Pearce1987,Alet2005,Malakis2009,Queiroz2011,Jin2012,Jin2013,Suzuki2015}, 
however; we demonstrate this at the finite-temperature deconfinement transition of a LGT for the first time. 
We perform extensive finite size scaling (FSS) studies on lattices upto $(512 a)^2$ ($a$ denotes the 
lattice spacing) with very small Trotter steps $\epsilon/J \sim 0.05$ ($J$ denotes a microscopic 
coupling) to demonstrate the BKT as well as the weak universality scenario at the deconfinement 
transition in the absence and presence of the $\pm 2$ charges, respectively. 

\begin{figure}
    \centering
    \includegraphics[scale=0.8]{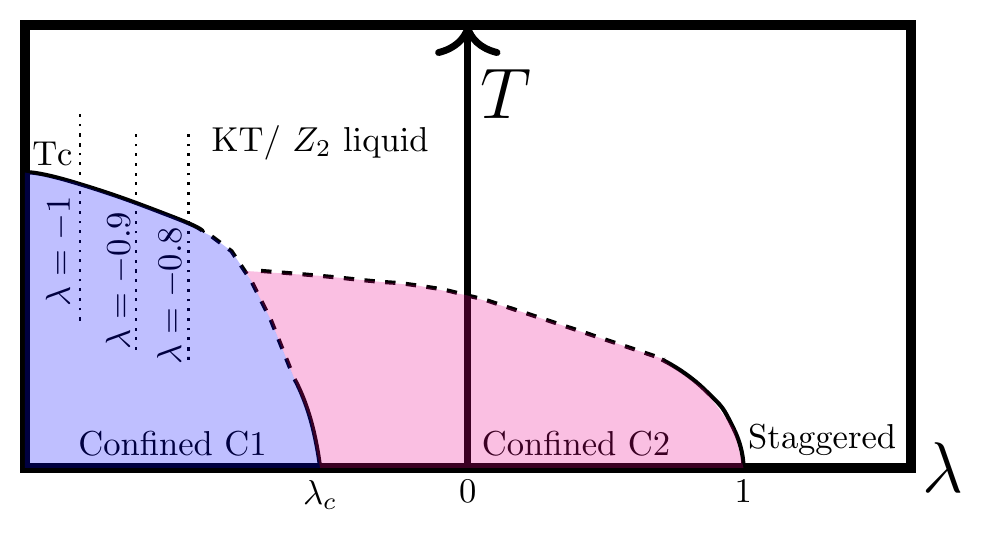}
    \caption{The $T-\lambda$ phase diagram: at $T=0$ there are two confined phases (C1 and C2) separated by
    a weak first order phase transition. Beyond the Rokshar-Kivelson point, $\lambda = 1$, the staggered phase
    is encountered. At high T, there is a pure $U(1)$ liquid, or a $Z_2$ liquid where $Q=\pm 2$ charges exist.
    In this paper, we study the finite temperature phase diagram. The dashed lines indicate a possible behaviour
    of the phase boundaries.}
    \label{fig:phDiag}
\end{figure}

\paragraph{Model, Simulations, and Phase Diagram.--} The quantum link gauge theories we consider use 
quantum spin $S = \frac{1}{2}$ as degrees of freedom on the bonds $(x,\h{\mu})$ between the sites of 
a square lattice. The electric flux operator, $E_{x,\h{\mu}} = S^3_{x,\h{\mu}}$, takes two values 
$\pm \frac{1}{2}$ while the gauge fields are the raising (lowering) operators of electric flux: 
$U_{x,\h{\mu}}^{(\dagger)} = S_{x,\h{\mu}}^{+(-)}$. The Hamiltonian operator is the sum of elementary 
plaquette terms 
\begin{equation}
H=-J\sum_\square \left[(U_\square + U_\square^\dagger)-\lambda(U_\square + U_\square^\dagger)^2 \right]
\label{eq:ham}
\end{equation}
where $U_\square = U_{x,\h{\mu}}U_{x+\h{\mu} ,\h{\nu}}U_{x+\h{\nu} ,\h{\mu}}^\dagger U_{x,\h{\nu}}^\dagger$. 
We set the lattice spacing of the square lattice ($a$) as well as $J$ to $1$ henceforth. \Cref{fig:setup} (a) 
shows the set-up of the lattice. The electric fluxes are shown as directed arrows. We note that 
$U_\square (U^\dagger_\square)$ reverses the orientation of the electric flux around the plaquette (clockwise
to anticlockwise and vice-versa), while all non-flippable plaquettes are annihilated. The $\lambda$ term is
akin to a potential energy, counting the total number of flippable plaquettes (both clockwise and anticlockwise).
The local $U(1)$ symmetry is generated by the Gauss Law operator 
$G_x=\sum_{\h{\mu}}  (E_{x,\h{\mu} }-E_{x-\h{\mu} ,\h{\mu}})$, and 
thus divides the Fock space into (exponentially) large number of superselection sectors. The physical states in 
the vacuum sector satisfy $G_x \ket{\psi}=0$ (six allowed states for each vertex of the square lattice) for all sites $x$. 
Additionally, the presence of $Q = \pm 2$, extends the Gauss Law to also include $G_x \ket{\phi}= \pm 2 \ket{\phi}$. 
Imposing periodic boundary conditions, only states which have zero total charge are allowed. \Cref{fig:setup} 
(b) shows all the allowed configurations (six with $Q_x=0$, one each with $Q_x=2$ and $Q_x = -2$) 
in our model, all allowed with equal weights. A detailed discussion about the symmetries is given in the 
Supp. Mat. At finite temperature $T = 1/\beta$, the equilibrium properties can be obtained from the partition 
function $Z$:
\begin{equation}
    Z = {\rm Tr} \left[ {\rm e}^{-\beta H}    
    \prod_x \{ 6\delta(G_x) + \delta(G_x - 2) + \delta(G_x + 2)\} \frac{1}{8}  
    \right],
    \label{eq:pf}
\end{equation}
 along with the charge neutrality constraint, $\sum_x Q_x = 0$.
 
  While the charges $Q = \pm 2$ do not have a kinetic energy term in the Hamiltonian, they 
are generated thermally, and move around the lattice due to thermal fluctuations. The $Q=\pm 2$ 
charges can be regarded as an example of \emph{annealed disorder} \cite{Cardy1996} where these 
``impurities" are in thermal equilibrium with the rest of the system according to \Cref{eq:pf}.
As one would naively expect from a confining theory, at $T=0$, the charges are high energy states, 
and thus do not appear. However, close to the deconfinement phase transition (when $T \sim \lambda J \sim M$, 
where $M$ is the rest mass of the charges), the charges are thermally generated, and therefore 
affect the critical properties of the system as we demonstrate later. In the Supp. Mat., we show 
how the presence of $Q = \pm 2$ charges gives rise to an effective $Z_2$ Gauss' Law for the theory.

\begin{figure}
    \includegraphics[width=9.5cm, height=6cm]{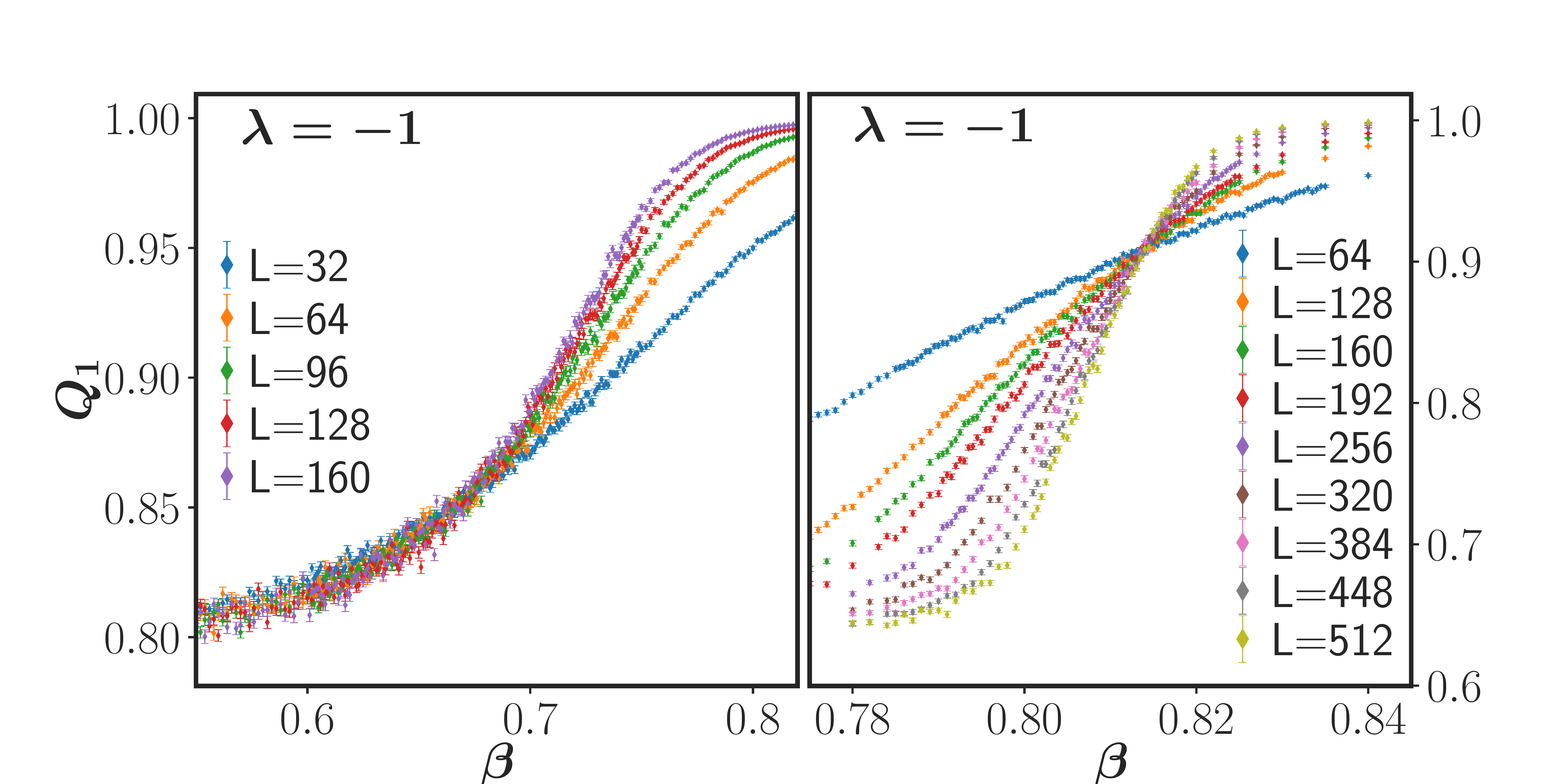}
    \caption{The Binder cumulant, $Q_1$, for the $U(1)$ QLM without (left) and with (right) the charges $Q=\pm 2$.
    The former displays an 2-d XY scenario (critical high-T and gapped low-T phase), while the latter exhibits a 
    2-d Ising scenario (gapped high-T and low-T) phases as one would expect from the SY analysis. We demonstrate
    weak universality for the latter case.}
    \label{fig:binderFlo}
\end{figure}

  In \Cref{fig:phDiag}, we sketch a finite temperature phase diagram of the model with and without the charges. The 
phase transition in $\lambda$ (at $T=0$) revealed two distinct crystalline confined phases (C1 and C2) separated by 
a weak first order phase transition at $\lambda_c \sim -0.36 $ \cite{Banerjee2013,Tschirsich2019}. The phases can 
be understood via a two-component magnetization (measured respectively on sublattice $A$ and $B$, see 
\Cref{fig:setup} (a)). In phase C1, both sublattices order (spontaneously breaking the lattice translation and 
charge conjugation), while for phase C2 only one of the sublattices order (breaking lattice translations). On 
raising the temperature, one expects the symmetry breakings to disappear accompanied by a spontaneous breaking of 
the global $U(1)^2$ symmetry, and SY analysis suggests a BKT phase transition. With the charges $Q = \pm 2$ in the 
ensemble, the quantum phase transition remains unchanged, while the thermal transition is modified. Due to the charges, 
the $U(1)^2$ centre symmetry is reduced to the $Z_2^2$ global center symmetry that breaks as the temperature is raised, 
leading one to naively expect a second order phase transition with 2-d $Z_2$ critical exponents. However, the thermal 
transition now displays properties associated with weak universality. 

 We study the model using a cluster quantum Monte Carlo (QMC) algorithm, which can efficiently update the 
(Kramers-Wannier) dualized version of the model \cite{Banerjee2021} on a lattice with $L$ ($L_T$) number 
of points in the spatial (temporal) direction. The algorithm builds clusters on the height variables 
$h^{A,B}$, placed at the centre of either the $A$ or the $B$ sublattices (see \Cref{fig:setup}), which are 
then flipped. A comparison of the QMC results, both in the absence and presence of $Q=\pm 2$ charges, with 
exact diagonalization (ED) results on small lattices is shown in Supp. Mat. The sublattice magnetizations 
are defined using the height variables as 
$M_{X} = \frac{1}{L_T}\sum_{\tilde{x}} \eta^X_{\tilde{x}} h^X_{\tilde{x}}$, where $X=A,B$ and $\tilde{x}$ 
denote the dual sites (centres of the plaquettes of the original space-time lattice). The phase factors 
$\eta^X_{\tilde{x}}$ are necessary to capture the ordering of the height variables corresponding to the 
flippability of the plaquettes. In the Supp. Mat., we show that $(M_{A}, M_{B})$ serve as order parameters for 
the deconfinement transition.  For finite size scaling (FSS) studies, we use total and connected susceptibilities:
\begin{equation}
 \chi_{\rm tot} = \frac{1}{V}\braket{M^2},~ 
 \chi_{\rm conn} = \frac{\beta}{V} \sum_{X}(\braket{M_X^2} - \braket{|M_X|}^2),
\end{equation}
 where $M^2 = \sum_{X=A,B} (M_X^2)$, $V = L^2$, and $\beta = \epsilon L_T$.
  Three different Binder cumulants were also used to estimate the critical exponents:
\begin{equation}
 \begin{split}
 Q_1 &= \frac{1}{2} \sum_X \frac{\braket{|M_X|}^2}{\braket{M_X^2}};~~
 Q_{2a} = 2 - \frac{\braket{M^4}}{\braket{M^2}^2};\\
 Q_{2b} &= \frac{3}{2} - \frac{1}{4} \sum_X \frac{\braket{M_X^4}}{\braket{M_X^2}^2}.
 \end{split}
\end{equation}

\begin{figure}
    \includegraphics[width=\hsize]{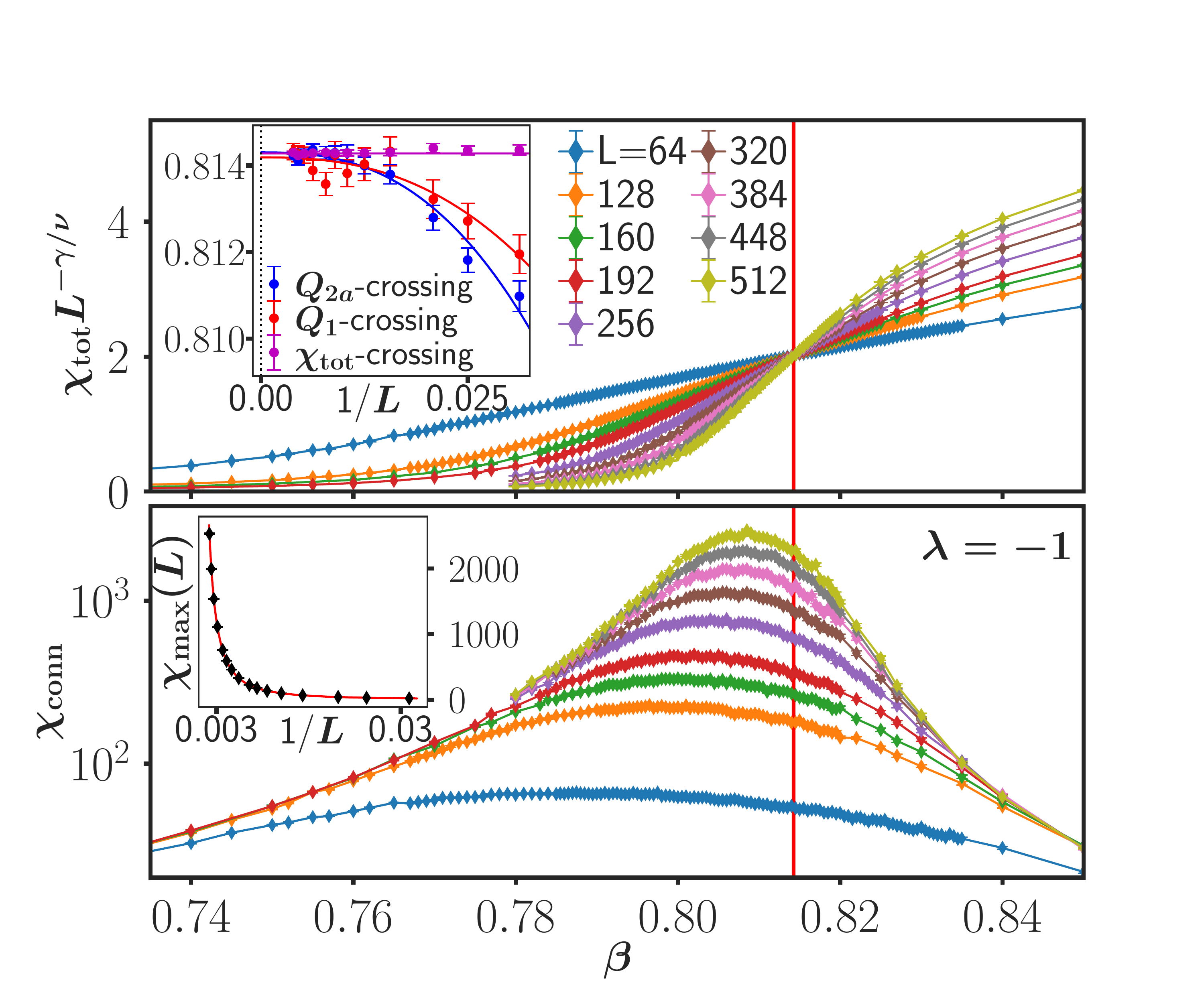}
    \caption{(Top) The critical temperature $\beta_c$ of the theory with charges is estimated by the
    crossing of $\chi_{\rm tot}$ for different $L$. (Inset) Estimates of $\beta_c$ using $Q_1$ and 
    $Q_{2a}$ yield consistent results. (Bottom) Plot of $\chi_{\rm conn}$ vs $\beta$ for various $L$
    shows a peak, whose scaling with L (inset) is used to extract $\gamma/\nu$. The vertical line
    denotes $\beta_c$.}
    \label{fig:susc}
\end{figure}

\paragraph{Finite temperature transition and FSS.--} The FSS hypothesis for the order parameter predicts 
that the Binder cumulants are universal at the critical point \cite{Binder1982,Beach2005,Wang2006}. 
We use this feature to show that the theory with and without the charges have different thermal behaviour. 
In the left panel of \Cref{fig:binderFlo} (the pure $U(1)$ theory), the Binder $Q_1$ curves for different 
lattice sizes $L$ collapse on each other at high-T, while they differ in the low-T phase. This is the expected 
behaviour for an XY-universality, where a critical phase goes into a massive phase through the BKT transition. 
The right panel shows $Q_1$ for the $U(1)$ theory with charges $Q = \pm 2$ for different $L$, which cross each 
other at $\beta_c \approx 0.815$, the probable location of a second order phase transition. We will postpone 
a detailed study of the BKT phase transition in a future publication, since it establishes the conventional wisdom, and 
concentrate here solely on the second order transition. 

  We extract the critical temperature $T_c = 1/\beta_c$ from crossing points of the curves of three different 
observables ($\chi_{\rm tot}L^{-\gamma/\nu}$, $Q_1$, and $Q_{2a}$). \Cref{fig:susc} (top panel) shows 
$\chi_{\rm tot} L^{-\gamma/\nu}$ vs $\beta$ for $L = 64, \cdots, 512$ and $L_T = 24$. In this analysis, we fixed 
$\gamma/\nu = 7/4$, the value for 2-d, $Z_2$ universality class. We will derive this independently later, and 
fitting for the ratio $\gamma/\nu$ only increases the uncertainties without any gain. Moreover, we need very 
precise estimates of $\beta_c$ to compute $\nu$. Using a second order polynomial interpolation to extract the 
crossing points of lattices $(L,2L)$, we observe a surprisingly flat behaviour for estimates of $\beta_c (L)$ 
from $\chi_{\rm tot}$ as a function of $1/L$, as shown in the inset. Estimates of $\beta_c$ were also extracted 
from the crossing points of $Q_1$ and $Q_{2a}$ curves. As shown in the inset, those estimates have larger finite 
size corrections, but yield the same $\beta_c$ for $L > 100$. Therefore, we quote the value of 
$\beta_c (L \rightarrow \infty, L_T=24)$ by fitting a constant to the $\beta_c$ estimates from $\chi_{\rm tot}$, 
and report it in \Cref{tab:CPs} for different $L_T$ values and three different $\lambda = -1.0, -0.9, -0.8$ values. 

\begin{table}
\begin{tabular}{ | c || c | c | c | c | c | }
 \hline
 $L_T$ & $\beta_c$    & $\eta$      & $\nu (Q_1)$ & $\nu (Q_{2a})$ & $\nu (Q_{2b})$\\
 \hline
 \multicolumn{6}{|c|}{$\lambda = -1.0$} \\
 \hline
 24 & 0.814279(14) & 0.2472(9)  & 1.35(2)  &  1.38(1)  &  1.38(2)  \\
 16 & 0.813783(15) & 0.2479(9) & 1.32(4)  &  1.34(2)  &  1.34(4)  \\
 8  & 0.811129(14) & 0.2489(8) & 1.33(3)  &  1.31(2)  &  1.34(3)  \\    
 4  & 0.801059(12) & 0.2509(8)  & 1.29(1)  &  1.31(1)  &  1.29(2)  \\    
 2  & 0.767685(10) & 0.2497(7)  & 1.19(1)  &  1.20(1)  &  1.20(1)  \\    
 \hline
 \multicolumn{6}{|c|}{$\lambda = -0.9$} \\
 \hline
 24    & 0.885292(17) & 0.2550 (18)& 1.45(3)  &  1.47(4) &  1.45(3)  \\
 \hline
 \multicolumn{6}{|c|}{$\lambda = -0.8$} \\
 \hline
 24    & 0.968196(26) & 0.2511 (10)& 1.64(9)  &  1.68(4)  &  1.64(8)  \\
 \hline
\end{tabular}
\caption{Estimates of $\beta_c$, $\eta$, and $\nu$ in the thermodynamic limit for 
different values of $L_T$ and $\lambda$.}
\label{tab:CPs}
\end{table}

 We turn to the estimate of the critical exponents. The scaling of the peak of $\chi_{\rm conn}$,
$\chi_{\rm conn, max} (L) = b L^{\gamma/\nu} = bL^{2-\eta}$ can be reliably used to extract $\eta$ (and thus
also $\gamma/\nu$). This quantity is shown (in a semi log scale) in \Cref{fig:susc} (bottom panel) 
vs $\beta$, with the vertical line indicating $\beta_c$ in the thermodynamic limit. The inset shows a 
power law fit to the $1/L$ dependence of $\chi_{\rm conn}$, from which $\eta$ is extracted, and reported for 
all our lattices and $\lambda$ values in \Cref{tab:CPs}. The systematic and the statistical 
errors are very well controlled in our data and analysis procedure, and yields estimates of $\eta$ for 
different $L_T$ and $\lambda$, consistent with the 2d $Z_2$ universality class. The largest deviation is 
only at the $3 \sigma$ level for $L_T=24$ and $\lambda = -1$, while most values are consistent with 
$\eta = \frac{1}{4}$ within $1 \sigma$. Moreover, this is also consistent with weak universality, where deviation
from the critical exponent ratios is not observed.

\begin{figure}
    \includegraphics[width=\hsize]{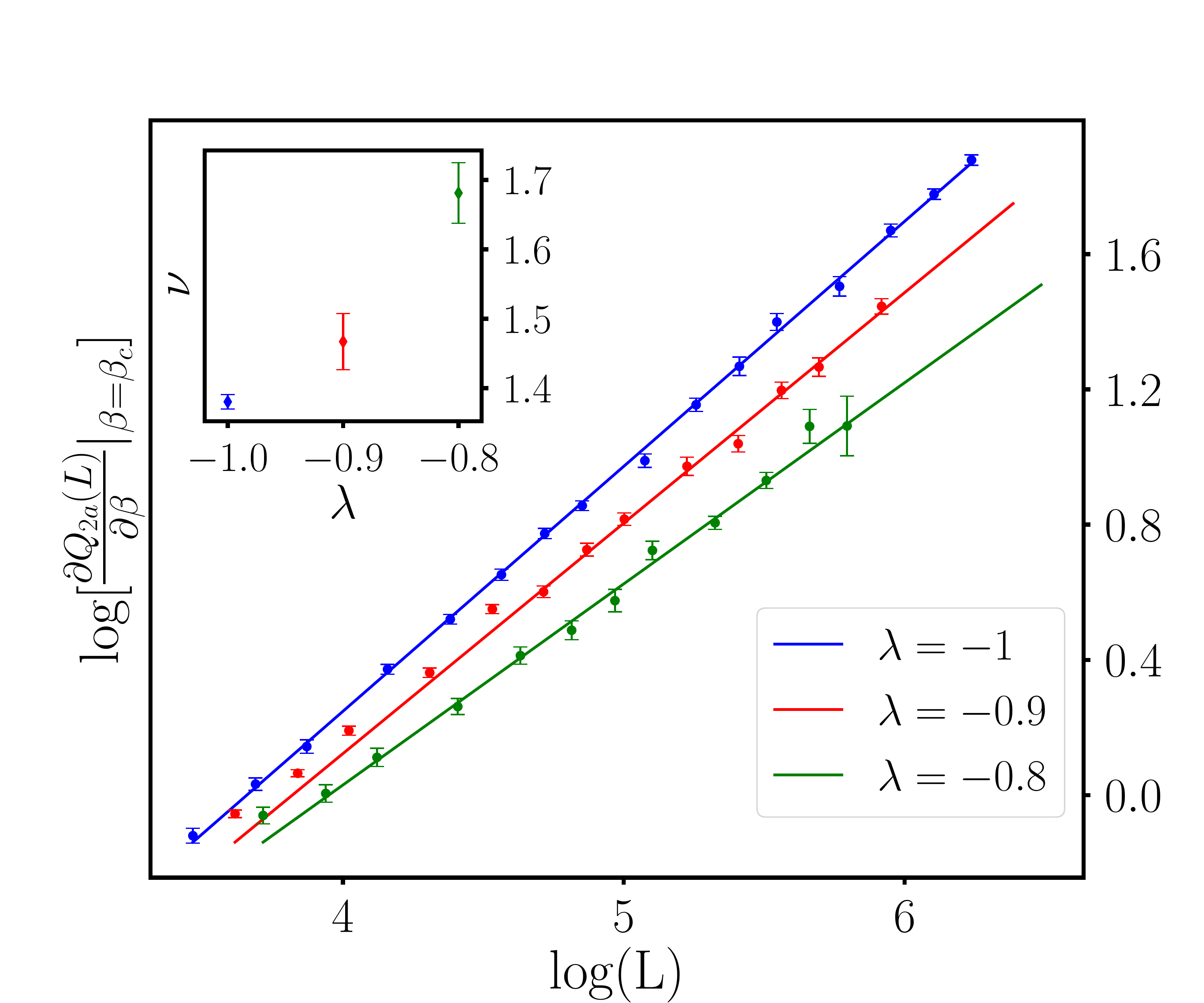}
    \caption{Extraction of $\nu$ from the slope of derivative of $Q_{2a}$ at $\beta_c$. The
    three curves correspond to different $\lambda$ which are vertically and horizontally 
    displaced for better visibility. The inset shows the change of slope (and hence $\nu$)
    with $\lambda$.}
    \label{fig:delQ2}
\end{figure}

 \paragraph{Weak universality.--} The next step to substantiate our claim involves the accurate computation 
of $\nu$ independently, for which we use all three Binder ratios. We employ the well-known result 
\cite{Wang2006,Hasenbusch2010} that for a dimensionless phenomenological coupling $R(\beta, L)$, the 
slope at $\beta_c$ directly yields the exponent $\nu$, 
$\left. \frac{\partial R (L)}{\partial \beta} \right|_{\beta_c} = a L^{1/\nu} (1 + bL^{-\omega})$.
For large lattices (or for large $\omega$), plotting the derivative (at $\beta_c$) vs $L$ in a 
log-log scale enables us to compute $1/\nu$ from the slope. The crucial requirement here is the very precise 
estimate of $\beta_c$, which we have already described. Performing this analysis using all three 
Binder ratios gives us consistent estimates of $\nu$. The particular analysis for $Q_{2a}$ for 
$L_T=24$ and three different $\lambda = -1.0, -0.9, -0.8$ is shown in \Cref{fig:delQ2}. To obtain 
the derivatives, we first fit the Binder ratios around $\beta_c$ to second order polynomials, and 
then take the derivative analytically with respect to $\beta$. Note that since the Binder ratios 
are all $O(1)$ numbers, and have a smooth behaviour around $\beta_c$, the polynomial fit is free 
of any systematic errors. Statistical errors are computed using bootstrapping samples from the 
entire data set. While the data is sufficiently accurate to extract reliable estimates of 
$\nu$, we are unable to estimate any reliable estimate of $\omega$, the leading correction to the 
scaling exponent. However, it is clear from the figure that the slopes of the curve (horizontally 
and vertically displaced for better visibility) are significantly different from the 2d $Z_2$ 
universality class value of $\nu = 1$. Instead, we witness significantly large values of $\nu$ as 
extracted from the coupling $Q_{2a}$: $1.38(1)$ for $\lambda = -1.0$, $1.47(4)$ for $\lambda = -0.9$, 
and $1.68(4)$ for $\lambda = -0.9$. These values are collected in \Cref{tab:CPs}, along with the 
corresponding estimates from the $Q_1$ and $Q_{2b}$. We note that all three estimates of $\nu$ at 
a fixed $\lambda$ agree with each other, and increase monotonically with $\lambda$. 
In particular, the inset of \Cref{fig:delQ2} displays this variation clearly. Not only are these 
values of $\nu$ anomalously large, but they also vary smoothly with the microscopic coupling. 
Both these features are hallmarks of weak universality. 

 An intriguing question is the reason for the occurrence of weak universality. According to the 
theory of renormalization group, a marginal operator \cite{Cardy1996, Jose1977, Delfino2019} 
is needed to generate a line of fixed points with continuously varying critical exponents. While 
we cannot yet offer an explicit realization of such an operator, we note that it would likely
depend on the $Q=\pm 2$ charges. In Supp. Mat., we show the behavior of $\braket{Q^2}$ (normalized 
with the volume) as a function of $\beta$ at $\lambda = -1, -0.9,-0.8$.
A small yet non-zero ($\braket{Q^2} \approx 0.03$ where $\braket{Q^2}_{\mathrm{max}}=1$ as 
$T\rightarrow \infty$) critical density of charges in the vicinity of $\beta_c$ suggests that 
charged operators play a non-trivial role in deciding the critical properties. 

\paragraph{Conclusions and outlook.--} In this article, we demonstrated that the presence of 
charges can completely alter the thermal phase transition in a pure gauge theory, using the example
of a $(2+1)$-d $U(1)$ QLM, by plotting the Binder cumulant across the phase
transition (\Cref{fig:binderFlo}). Using $\chi_{\rm tot}(L)$, and the Binder ratios, we located 
$\beta_c$ very accurately, and then used the scaling of $\chi_{\rm conn, max}(L)$ to 
compute $\eta$, which is compatible with the 2-d $Z_2$ value, as expected from 
universality arguments. Finally, using three Binder ratios, we demonstrate that the individual 
exponents $\nu$ (and $\gamma$) have anomalously large values compared to the 2-d $Z_2$ value, and 
furthermore vary smoothly as a function of the microscopic coupling $\lambda$. These three pieces of 
evidence conclusively show that weak universality is relevant for this thermal 
transition instead of the usual universality scenario. 

 Our results open some very intriguing directions for further research. A close examination of the 
charge and the electric flux distribution at $\beta_c$ could help to provide a better understanding 
of a possible marginal operator that leads to weak universality. It is also interesting to follow the 
thermal transition to more negative values of $\lambda$ to explore whether it reaches the 2d $Z_2$ limit. 
Finally, exploring the phase diagram where three phases (C1, C2, and KT/$Z_2$ liquid) meet is an 
exciting project for the future. The prospect that this model could be realized in quantum simulator 
set-ups in the recent future makes these results potentially interesting \cite{Marcos2014,Celi2019}.
Whether such weak universality also happens for other gauge theories is an open question to be explored 
by the community.

 We would like to thank Shailesh Chandrasekharan and Uwe-Jens Wiese for useful discussions. 
 We acknowledge the computational resources provided by IACS and SINP. 
D.B. acknowledges assistance from SERB Starting grant SRG/2021/000396-C from the DST (Govt. of India). 

\bibliography{letter.bib}

\begin{thebibliography}{40}%
\makeatletter
\providecommand \@ifxundefined [1]{%
 \@ifx{#1\undefined}
}%
\providecommand \@ifnum [1]{%
 \ifnum #1\expandafter \@firstoftwo
 \else \expandafter \@secondoftwo
 \fi
}%
\providecommand \@ifx [1]{%
 \ifx #1\expandafter \@firstoftwo
 \else \expandafter \@secondoftwo
 \fi
}%
\providecommand \natexlab [1]{#1}%
\providecommand \enquote  [1]{``#1''}%
\providecommand \bibnamefont  [1]{#1}%
\providecommand \bibfnamefont [1]{#1}%
\providecommand \citenamefont [1]{#1}%
\providecommand \href@noop [0]{\@secondoftwo}%
\providecommand \href [0]{\begingroup \@sanitize@url \@href}%
\providecommand \@href[1]{\@@startlink{#1}\@@href}%
\providecommand \@@href[1]{\endgroup#1\@@endlink}%
\providecommand \@sanitize@url [0]{\catcode `\\12\catcode `\$12\catcode
  `\&12\catcode `\#12\catcode `\^12\catcode `\_12\catcode `\%12\relax}%
\providecommand \@@startlink[1]{}%
\providecommand \@@endlink[0]{}%
\providecommand \url  [0]{\begingroup\@sanitize@url \@url }%
\providecommand \@url [1]{\endgroup\@href {#1}{\urlprefix }}%
\providecommand \urlprefix  [0]{URL }%
\providecommand \Eprint [0]{\href }%
\providecommand \doibase [0]{http://dx.doi.org/}%
\providecommand \selectlanguage [0]{\@gobble}%
\providecommand \bibinfo  [0]{\@secondoftwo}%
\providecommand \bibfield  [0]{\@secondoftwo}%
\providecommand \translation [1]{[#1]}%
\providecommand \BibitemOpen [0]{}%
\providecommand \bibitemStop [0]{}%
\providecommand \bibitemNoStop [0]{.\EOS\space}%
\providecommand \EOS [0]{\spacefactor3000\relax}%
\providecommand \BibitemShut  [1]{\csname bibitem#1\endcsname}%
\let\auto@bib@innerbib\@empty
\bibitem [{\citenamefont {Wilczek}(1999)}]{Wilczek1999}%
  \BibitemOpen
  \bibfield  {author} {\bibinfo {author} {\bibfnamefont {Frank}\ \bibnamefont
  {Wilczek}},\ }\bibfield  {title} {\enquote {\bibinfo {title} {{QCD in extreme
  conditions}},}\ }\bibfield  {booktitle} {\emph {\bibinfo {booktitle} {{9th
  CRM Summer School: Theoretical Physics at the End of the 20th Century}}},\
  }\href@noop {} {\ ,\ \bibinfo {pages} {567--636} (\bibinfo {year} {1999})},\
  \Eprint {http://arxiv.org/abs/hep-ph/0003183} {arXiv:hep-ph/0003183}
  \BibitemShut {NoStop}%
\bibitem [{\citenamefont {Philipsen}(2019)}]{Philipsen2019}%
  \BibitemOpen
  \bibfield  {author} {\bibinfo {author} {\bibfnamefont {Owe}\ \bibnamefont
  {Philipsen}},\ }\bibfield  {title} {\enquote {\bibinfo {title} {{Constraining
  the phase diagram of QCD at finite temperature and density}},}\ }\href
  {\doibase 10.22323/1.363.0273} {\bibfield  {journal} {\bibinfo  {journal}
  {PoS}\ }\textbf {\bibinfo {volume} {LATTICE2019}},\ \bibinfo {pages} {273}
  (\bibinfo {year} {2019})},\ \Eprint {http://arxiv.org/abs/1912.04827}
  {arXiv:1912.04827 [hep-lat]} \BibitemShut {NoStop}%
\bibitem [{\citenamefont {D'Elia}(2019)}]{DElia2018}%
  \BibitemOpen
  \bibfield  {author} {\bibinfo {author} {\bibfnamefont {Massimo}\ \bibnamefont
  {D'Elia}},\ }\bibfield  {title} {\enquote {\bibinfo {title}
  {{High-Temperature QCD: theory overview}},}\ }\href {\doibase
  10.1016/j.nuclphysa.2018.10.042} {\bibfield  {journal} {\bibinfo  {journal}
  {Nucl. Phys. A}\ }\textbf {\bibinfo {volume} {982}},\ \bibinfo {pages}
  {99--105} (\bibinfo {year} {2019})},\ \Eprint
  {http://arxiv.org/abs/1809.10660} {arXiv:1809.10660 [hep-lat]} \BibitemShut
  {NoStop}%
\bibitem [{\citenamefont {Sharma}(2021)}]{Sharma2021}%
  \BibitemOpen
  \bibfield  {author} {\bibinfo {author} {\bibfnamefont {Sayantan}\
  \bibnamefont {Sharma}},\ }\bibfield  {title} {\enquote {\bibinfo {title}
  {{Recent theoretical developments on QCD matter at finite temperature and
  density}},}\ }\href {\doibase 10.1142/S0218301321300034} {\bibfield
  {journal} {\bibinfo  {journal} {Int. J. Mod. Phys. E}\ }\textbf {\bibinfo
  {volume} {30}},\ \bibinfo {pages} {2130003} (\bibinfo {year} {2021})},\
  \Eprint {http://arxiv.org/abs/2103.13641} {arXiv:2103.13641 [hep-lat]}
  \BibitemShut {NoStop}%
\bibitem [{\citenamefont {Kronfeld}\ \emph {et~al.}(2022)\citenamefont
  {Kronfeld} \emph {et~al.}}]{USQCD2022}%
  \BibitemOpen
  \bibfield  {author} {\bibinfo {author} {\bibfnamefont {Andreas~S.}\
  \bibnamefont {Kronfeld}} \emph {et~al.} (\bibinfo {collaboration} {USQCD}),\
  }\bibfield  {title} {\enquote {\bibinfo {title} {{Lattice QCD and Particle
  Physics}},}\ }\href@noop {} {\  (\bibinfo {year} {2022})},\ \Eprint
  {http://arxiv.org/abs/2207.07641} {arXiv:2207.07641 [hep-lat]} \BibitemShut
  {NoStop}%
\bibitem [{\citenamefont {Creutz}\ and\ \citenamefont
  {Moriarty}(1982)}]{Creutz1982}%
  \BibitemOpen
  \bibfield  {author} {\bibinfo {author} {\bibfnamefont {Michael}\ \bibnamefont
  {Creutz}}\ and\ \bibinfo {author} {\bibfnamefont {K.~J.~M.}\ \bibnamefont
  {Moriarty}},\ }\bibfield  {title} {\enquote {\bibinfo {title} {{Numerical
  Studies of Wilson Loops in SU(3) Gauge Theory in Four-dimensions}},}\ }\href
  {\doibase 10.1103/PhysRevD.26.2166} {\bibfield  {journal} {\bibinfo
  {journal} {Phys. Rev. D}\ }\textbf {\bibinfo {volume} {26}},\ \bibinfo
  {pages} {2166} (\bibinfo {year} {1982})}\BibitemShut {NoStop}%
\bibitem [{\citenamefont {Kogut}\ \emph {et~al.}(1985)\citenamefont {Kogut},
  \citenamefont {Polonyi}, \citenamefont {Wyld}, \citenamefont {Shigemitsu},\
  and\ \citenamefont {Sinclair}}]{Kogut1984}%
  \BibitemOpen
  \bibfield  {author} {\bibinfo {author} {\bibfnamefont {John~B.}\ \bibnamefont
  {Kogut}}, \bibinfo {author} {\bibfnamefont {J.}~\bibnamefont {Polonyi}},
  \bibinfo {author} {\bibfnamefont {H.~W.}\ \bibnamefont {Wyld}}, \bibinfo
  {author} {\bibfnamefont {J.}~\bibnamefont {Shigemitsu}}, \ and\ \bibinfo
  {author} {\bibfnamefont {D.~K.}\ \bibnamefont {Sinclair}},\ }\bibfield
  {title} {\enquote {\bibinfo {title} {{Further Evidence for the First Order
  Nature of the Pure Gauge SU(3) Deconfinement Transition}},}\ }\href {\doibase
  10.1016/0550-3213(85)90264-0} {\bibfield  {journal} {\bibinfo  {journal}
  {Nucl. Phys. B}\ }\textbf {\bibinfo {volume} {251}},\ \bibinfo {pages}
  {311--332} (\bibinfo {year} {1985})}\BibitemShut {NoStop}%
\bibitem [{\citenamefont {Creutz}(1980)}]{Creutz1980}%
  \BibitemOpen
  \bibfield  {author} {\bibinfo {author} {\bibfnamefont {M.}~\bibnamefont
  {Creutz}},\ }\bibfield  {title} {\enquote {\bibinfo {title} {{Monte Carlo
  Study of Quantized SU(2) Gauge Theory}},}\ }\href {\doibase
  10.1103/PhysRevD.21.2308} {\bibfield  {journal} {\bibinfo  {journal} {Phys.
  Rev. D}\ }\textbf {\bibinfo {volume} {21}},\ \bibinfo {pages} {2308--2315}
  (\bibinfo {year} {1980})}\BibitemShut {NoStop}%
\bibitem [{\citenamefont {Gavai}(1983)}]{Gavai1982}%
  \BibitemOpen
  \bibfield  {author} {\bibinfo {author} {\bibfnamefont {R.~V.}\ \bibnamefont
  {Gavai}},\ }\bibfield  {title} {\enquote {\bibinfo {title} {{The
  Deconfinement Transition in SU(2) Lattice Gauge Theories}},}\ }\href
  {\doibase 10.1016/0550-3213(83)90255-9} {\bibfield  {journal} {\bibinfo
  {journal} {Nucl. Phys. B}\ }\textbf {\bibinfo {volume} {215}},\ \bibinfo
  {pages} {458--469} (\bibinfo {year} {1983})}\BibitemShut {NoStop}%
\bibitem [{\citenamefont {Hasenfratz}\ \emph {et~al.}(1983)\citenamefont
  {Hasenfratz}, \citenamefont {Karsch},\ and\ \citenamefont
  {Stamatescu}}]{Hasenfratz1983}%
  \BibitemOpen
  \bibfield  {author} {\bibinfo {author} {\bibfnamefont {P.}~\bibnamefont
  {Hasenfratz}}, \bibinfo {author} {\bibfnamefont {F.}~\bibnamefont {Karsch}},
  \ and\ \bibinfo {author} {\bibfnamefont {I.~O.}\ \bibnamefont {Stamatescu}},\
  }\bibfield  {title} {\enquote {\bibinfo {title} {{The SU(3) Deconfinement
  Phase Transition in the Presence of Quarks}},}\ }\href {\doibase
  10.1016/0370-2693(83)90565-8} {\bibfield  {journal} {\bibinfo  {journal}
  {Phys. Lett. B}\ }\textbf {\bibinfo {volume} {133}},\ \bibinfo {pages}
  {221--226} (\bibinfo {year} {1983})}\BibitemShut {NoStop}%
\bibitem [{\citenamefont {Bhattacharya}\ \emph {et~al.}(2014)\citenamefont
  {Bhattacharya} \emph {et~al.}}]{Bhattacharya2014}%
  \BibitemOpen
  \bibfield  {author} {\bibinfo {author} {\bibfnamefont {Tanmoy}\ \bibnamefont
  {Bhattacharya}} \emph {et~al.},\ }\bibfield  {title} {\enquote {\bibinfo
  {title} {{QCD Phase Transition with Chiral Quarks and Physical Quark
  Masses}},}\ }\href {\doibase 10.1103/PhysRevLett.113.082001} {\bibfield
  {journal} {\bibinfo  {journal} {Phys. Rev. Lett.}\ }\textbf {\bibinfo
  {volume} {113}},\ \bibinfo {pages} {082001} (\bibinfo {year} {2014})},\
  \Eprint {http://arxiv.org/abs/1402.5175} {arXiv:1402.5175 [hep-lat]}
  \BibitemShut {NoStop}%
\bibitem [{\citenamefont {{Svetitsky}}\ and\ \citenamefont
  {{Yaffe}}(1982)}]{Svetitsky1982}%
  \BibitemOpen
  \bibfield  {author} {\bibinfo {author} {\bibfnamefont {Benjamin}\
  \bibnamefont {{Svetitsky}}}\ and\ \bibinfo {author} {\bibfnamefont
  {Laurence~G.}\ \bibnamefont {{Yaffe}}},\ }\bibfield  {title} {\enquote
  {\bibinfo {title} {{Critical behavior at finite-temperature confinement
  transitions}},}\ }\href {\doibase 10.1016/0550-3213(82)90172-9} {\bibfield
  {journal} {\bibinfo  {journal} {Nuclear Physics B}\ }\textbf {\bibinfo
  {volume} {210}},\ \bibinfo {pages} {423--447} (\bibinfo {year}
  {1982})}\BibitemShut {NoStop}%
\bibitem [{\citenamefont {Caselle}\ and\ \citenamefont
  {Hasenbusch}(1996)}]{Caselle1995}%
  \BibitemOpen
  \bibfield  {author} {\bibinfo {author} {\bibfnamefont {M.}~\bibnamefont
  {Caselle}}\ and\ \bibinfo {author} {\bibfnamefont {M.}~\bibnamefont
  {Hasenbusch}},\ }\bibfield  {title} {\enquote {\bibinfo {title}
  {{Deconfinement transition and dimensional crossover in the 3-D gauge Ising
  model}},}\ }\href {\doibase 10.1016/0550-3213(96)00161-7} {\bibfield
  {journal} {\bibinfo  {journal} {Nucl. Phys. B}\ }\textbf {\bibinfo {volume}
  {470}},\ \bibinfo {pages} {435--453} (\bibinfo {year} {1996})},\ \Eprint
  {http://arxiv.org/abs/hep-lat/9511015} {arXiv:hep-lat/9511015} \BibitemShut
  {NoStop}%
\bibitem [{\citenamefont {Bonati}\ and\ \citenamefont
  {D'Elia}(2013)}]{Bonati2013}%
  \BibitemOpen
  \bibfield  {author} {\bibinfo {author} {\bibfnamefont {Claudio}\ \bibnamefont
  {Bonati}}\ and\ \bibinfo {author} {\bibfnamefont {Massimo}\ \bibnamefont
  {D'Elia}},\ }\bibfield  {title} {\enquote {\bibinfo {title} {{Phase diagram
  of the 4D U(1) model at finite temperature}},}\ }\href {\doibase
  10.1103/PhysRevD.88.065025} {\bibfield  {journal} {\bibinfo  {journal} {Phys.
  Rev. D}\ }\textbf {\bibinfo {volume} {88}},\ \bibinfo {pages} {065025}
  (\bibinfo {year} {2013})},\ \Eprint {http://arxiv.org/abs/1305.3564}
  {arXiv:1305.3564 [hep-lat]} \BibitemShut {NoStop}%
\bibitem [{\citenamefont {Lau}\ and\ \citenamefont {Teper}(2016)}]{Lau2015}%
  \BibitemOpen
  \bibfield  {author} {\bibinfo {author} {\bibfnamefont {Richard}\ \bibnamefont
  {Lau}}\ and\ \bibinfo {author} {\bibfnamefont {Michael}\ \bibnamefont
  {Teper}},\ }\bibfield  {title} {\enquote {\bibinfo {title} {{The deconfining
  phase transition of SO(N) gauge theories in 2+1 dimensions}},}\ }\href
  {\doibase 10.1007/JHEP03(2016)072} {\bibfield  {journal} {\bibinfo  {journal}
  {JHEP}\ }\textbf {\bibinfo {volume} {03}},\ \bibinfo {pages} {072} (\bibinfo
  {year} {2016})},\ \Eprint {http://arxiv.org/abs/1510.07841} {arXiv:1510.07841
  [hep-lat]} \BibitemShut {NoStop}%
\bibitem [{\citenamefont {Borisenko}\ \emph {et~al.}(2016)\citenamefont
  {Borisenko}, \citenamefont {Chelnokov}, \citenamefont {Cuteri},\ and\
  \citenamefont {Papa}}]{Borisenko2015}%
  \BibitemOpen
  \bibfield  {author} {\bibinfo {author} {\bibfnamefont {Oleg}\ \bibnamefont
  {Borisenko}}, \bibinfo {author} {\bibfnamefont {Volodymyr}\ \bibnamefont
  {Chelnokov}}, \bibinfo {author} {\bibfnamefont {Francesca}\ \bibnamefont
  {Cuteri}}, \ and\ \bibinfo {author} {\bibfnamefont {Alessandro}\ \bibnamefont
  {Papa}},\ }\bibfield  {title} {\enquote {\bibinfo {title}
  {{Berezinskii-Kosterlitz-Thouless phase transitions in two-dimensional
  non-Abelian spin models}},}\ }\href {\doibase 10.1103/PhysRevE.94.012108}
  {\bibfield  {journal} {\bibinfo  {journal} {Phys. Rev. E}\ }\textbf {\bibinfo
  {volume} {94}},\ \bibinfo {pages} {012108} (\bibinfo {year} {2016})},\
  \Eprint {http://arxiv.org/abs/1512.05737} {arXiv:1512.05737 [hep-lat]}
  \BibitemShut {NoStop}%
\bibitem [{\citenamefont {Biswal}\ \emph {et~al.}(2017)\citenamefont {Biswal},
  \citenamefont {Deka}, \citenamefont {Digal},\ and\ \citenamefont
  {Saumia}}]{Biswal2016}%
  \BibitemOpen
  \bibfield  {author} {\bibinfo {author} {\bibfnamefont {Minati}\ \bibnamefont
  {Biswal}}, \bibinfo {author} {\bibfnamefont {Mridupawan}\ \bibnamefont
  {Deka}}, \bibinfo {author} {\bibfnamefont {Sanatan}\ \bibnamefont {Digal}}, \
  and\ \bibinfo {author} {\bibfnamefont {P.~S.}\ \bibnamefont {Saumia}},\
  }\bibfield  {title} {\enquote {\bibinfo {title} {{Confinement-deconfinement
  transition in $SU(2)$ Higgs theory}},}\ }\href {\doibase
  10.1103/PhysRevD.96.014503} {\bibfield  {journal} {\bibinfo  {journal} {Phys.
  Rev. D}\ }\textbf {\bibinfo {volume} {96}},\ \bibinfo {pages} {014503}
  (\bibinfo {year} {2017})},\ \Eprint {http://arxiv.org/abs/1610.08265}
  {arXiv:1610.08265 [hep-lat]} \BibitemShut {NoStop}%
\bibitem [{\citenamefont {Kuramashi}\ and\ \citenamefont
  {Yoshimura}(2019)}]{Kuramashi2018}%
  \BibitemOpen
  \bibfield  {author} {\bibinfo {author} {\bibfnamefont {Yoshinobu}\
  \bibnamefont {Kuramashi}}\ and\ \bibinfo {author} {\bibfnamefont {Yusuke}\
  \bibnamefont {Yoshimura}},\ }\bibfield  {title} {\enquote {\bibinfo {title}
  {{Three-dimensional finite temperature Z$_{2}$ gauge theory with tensor
  network scheme}},}\ }\href {\doibase 10.1007/JHEP08(2019)023} {\bibfield
  {journal} {\bibinfo  {journal} {JHEP}\ }\textbf {\bibinfo {volume} {08}},\
  \bibinfo {pages} {023} (\bibinfo {year} {2019})},\ \Eprint
  {http://arxiv.org/abs/1808.08025} {arXiv:1808.08025 [hep-lat]} \BibitemShut
  {NoStop}%
\bibitem [{\citenamefont {{Suzuki}}(1974)}]{Suzuki1974}%
  \BibitemOpen
  \bibfield  {author} {\bibinfo {author} {\bibfnamefont {M.}~\bibnamefont
  {{Suzuki}}},\ }\bibfield  {title} {\enquote {\bibinfo {title} {{New
  Universality of Critical Exponents}},}\ }\href {\doibase 10.1143/PTP.51.1992}
  {\bibfield  {journal} {\bibinfo  {journal} {Progress of Theoretical Physics}\
  }\textbf {\bibinfo {volume} {51}},\ \bibinfo {pages} {1992--1993} (\bibinfo
  {year} {1974})}\BibitemShut {NoStop}%
\bibitem [{\citenamefont {{Baxter}}(1971)}]{Baxter1971}%
  \BibitemOpen
  \bibfield  {author} {\bibinfo {author} {\bibfnamefont {R.~J.}\ \bibnamefont
  {{Baxter}}},\ }\bibfield  {title} {\enquote {\bibinfo {title} {{Eight-Vertex
  Model in Lattice Statistics}},}\ }\href {\doibase 10.1103/PhysRevLett.26.832}
  {\bibfield  {journal} {\bibinfo  {journal} {Physical Review Letters}\
  }\textbf {\bibinfo {volume} {26}},\ \bibinfo {pages} {832--833} (\bibinfo
  {year} {1971})}\BibitemShut {NoStop}%
\bibitem [{\citenamefont {{Ashkin}}\ and\ \citenamefont
  {{Teller}}(1943)}]{Ashkin1943}%
  \BibitemOpen
  \bibfield  {author} {\bibinfo {author} {\bibfnamefont {J.}~\bibnamefont
  {{Ashkin}}}\ and\ \bibinfo {author} {\bibfnamefont {E.}~\bibnamefont
  {{Teller}}},\ }\bibfield  {title} {\enquote {\bibinfo {title} {{Statistics of
  Two-Dimensional Lattices with Four Components}},}\ }\href {\doibase
  10.1103/PhysRev.64.178} {\bibfield  {journal} {\bibinfo  {journal} {Physical
  Review}\ }\textbf {\bibinfo {volume} {64}},\ \bibinfo {pages} {178--184}
  (\bibinfo {year} {1943})}\BibitemShut {NoStop}%
\bibitem [{\citenamefont {{Pearce}}\ and\ \citenamefont
  {{Kim}}(1987)}]{Pearce1987}%
  \BibitemOpen
  \bibfield  {author} {\bibinfo {author} {\bibfnamefont {P.~A.}\ \bibnamefont
  {{Pearce}}}\ and\ \bibinfo {author} {\bibfnamefont {D.}~\bibnamefont
  {{Kim}}},\ }\bibfield  {title} {\enquote {\bibinfo {title} {{Continuously
  varying exponents in magnetic hard squares}},}\ }\href {\doibase
  10.1088/0305-4470/20/18/044} {\bibfield  {journal} {\bibinfo  {journal}
  {Journal of Physics A Mathematical General}\ }\textbf {\bibinfo {volume}
  {20}},\ \bibinfo {pages} {6471--6485} (\bibinfo {year} {1987})}\BibitemShut
  {NoStop}%
\bibitem [{\citenamefont {{Alet}}\ \emph {et~al.}(2005)\citenamefont {{Alet}},
  \citenamefont {{Jacobsen}}, \citenamefont {{Misguich}}, \citenamefont
  {{Pasquier}}, \citenamefont {{Mila}},\ and\ \citenamefont
  {{Troyer}}}]{Alet2005}%
  \BibitemOpen
  \bibfield  {author} {\bibinfo {author} {\bibfnamefont {Fabien}\ \bibnamefont
  {{Alet}}}, \bibinfo {author} {\bibfnamefont {Jesper~Lykke}\ \bibnamefont
  {{Jacobsen}}}, \bibinfo {author} {\bibfnamefont {Gr{\'e}goire}\ \bibnamefont
  {{Misguich}}}, \bibinfo {author} {\bibfnamefont {Vincent}\ \bibnamefont
  {{Pasquier}}}, \bibinfo {author} {\bibfnamefont {Fr{\'e}d{\'e}ric}\
  \bibnamefont {{Mila}}}, \ and\ \bibinfo {author} {\bibfnamefont {Matthias}\
  \bibnamefont {{Troyer}}},\ }\bibfield  {title} {\enquote {\bibinfo {title}
  {{Interacting Classical Dimers on the Square Lattice}},}\ }\href {\doibase
  10.1103/PhysRevLett.94.235702} {\bibfield  {journal} {\bibinfo  {journal}
  {Physical Review Letters}\ }\textbf {\bibinfo {volume} {94}},\ \bibinfo {eid}
  {235702} (\bibinfo {year} {2005})},\ \Eprint
  {http://arxiv.org/abs/cond-mat/0501241} {arXiv:cond-mat/0501241
  [cond-mat.stat-mech]} \BibitemShut {NoStop}%
\bibitem [{\citenamefont {{Malakis}}\ \emph {et~al.}(2009)\citenamefont
  {{Malakis}}, \citenamefont {{Berker}}, \citenamefont {{Hadjiagapiou}},\ and\
  \citenamefont {{Fytas}}}]{Malakis2009}%
  \BibitemOpen
  \bibfield  {author} {\bibinfo {author} {\bibfnamefont {A.}~\bibnamefont
  {{Malakis}}}, \bibinfo {author} {\bibfnamefont {A.~Nihat}\ \bibnamefont
  {{Berker}}}, \bibinfo {author} {\bibfnamefont {I.~A.}\ \bibnamefont
  {{Hadjiagapiou}}}, \ and\ \bibinfo {author} {\bibfnamefont {N.~G.}\
  \bibnamefont {{Fytas}}},\ }\bibfield  {title} {\enquote {\bibinfo {title}
  {{Strong violation of critical phenomena universality: Wang-Landau study of
  the two-dimensional Blume-Capel model under bond randomness}},}\ }\href
  {\doibase 10.1103/PhysRevE.79.011125} {\bibfield  {journal} {\bibinfo
  {journal} {Physical Review E}\ }\textbf {\bibinfo {volume} {79}},\ \bibinfo
  {eid} {011125} (\bibinfo {year} {2009})},\ \Eprint
  {http://arxiv.org/abs/0809.4241} {arXiv:0809.4241 [cond-mat.dis-nn]}
  \BibitemShut {NoStop}%
\bibitem [{\citenamefont {{de Queiroz}}(2011)}]{Queiroz2011}%
  \BibitemOpen
  \bibfield  {author} {\bibinfo {author} {\bibfnamefont {S.~L.~A.}\
  \bibnamefont {{de Queiroz}}},\ }\bibfield  {title} {\enquote {\bibinfo
  {title} {{Scaling behavior of a square-lattice Ising model with competing
  interactions in a uniform field}},}\ }\href {\doibase
  10.1103/PhysRevE.84.031132} {\bibfield  {journal} {\bibinfo  {journal}
  {Physical Review E}\ }\textbf {\bibinfo {volume} {84}},\ \bibinfo {eid}
  {031132} (\bibinfo {year} {2011})},\ \Eprint {http://arxiv.org/abs/1107.6022}
  {arXiv:1107.6022 [cond-mat.stat-mech]} \BibitemShut {NoStop}%
\bibitem [{\citenamefont {{Jin}}\ \emph {et~al.}(2012)\citenamefont {{Jin}},
  \citenamefont {{Sen}},\ and\ \citenamefont {{Sandvik}}}]{Jin2012}%
  \BibitemOpen
  \bibfield  {author} {\bibinfo {author} {\bibfnamefont {Songbo}\ \bibnamefont
  {{Jin}}}, \bibinfo {author} {\bibfnamefont {Arnab}\ \bibnamefont {{Sen}}}, \
  and\ \bibinfo {author} {\bibfnamefont {Anders~W.}\ \bibnamefont
  {{Sandvik}}},\ }\bibfield  {title} {\enquote {\bibinfo {title}
  {{Ashkin-Teller Criticality and Pseudo-First-Order Behavior in a Frustrated
  Ising Model on the Square Lattice}},}\ }\href {\doibase
  10.1103/PhysRevLett.108.045702} {\bibfield  {journal} {\bibinfo  {journal}
  {Physical Review Letters}\ }\textbf {\bibinfo {volume} {108}},\ \bibinfo
  {eid} {045702} (\bibinfo {year} {2012})},\ \Eprint
  {http://arxiv.org/abs/1110.5874} {arXiv:1110.5874 [cond-mat.stat-mech]}
  \BibitemShut {NoStop}%
\bibitem [{\citenamefont {Jin}\ \emph {et~al.}(2013)\citenamefont {Jin},
  \citenamefont {Sen}, \citenamefont {Guo},\ and\ \citenamefont
  {Sandvik}}]{Jin2013}%
  \BibitemOpen
  \bibfield  {author} {\bibinfo {author} {\bibfnamefont {Songbo}\ \bibnamefont
  {Jin}}, \bibinfo {author} {\bibfnamefont {Arnab}\ \bibnamefont {Sen}},
  \bibinfo {author} {\bibfnamefont {Wenan}\ \bibnamefont {Guo}}, \ and\
  \bibinfo {author} {\bibfnamefont {Anders~W.}\ \bibnamefont {Sandvik}},\
  }\bibfield  {title} {\enquote {\bibinfo {title} {Phase transitions in the
  frustrated ising model on the square lattice},}\ }\href {\doibase
  10.1103/PhysRevB.87.144406} {\bibfield  {journal} {\bibinfo  {journal} {Phys.
  Rev. B}\ }\textbf {\bibinfo {volume} {87}},\ \bibinfo {pages} {144406}
  (\bibinfo {year} {2013})}\BibitemShut {NoStop}%
\bibitem [{\citenamefont {{Suzuki}}\ \emph {et~al.}(2015)\citenamefont
  {{Suzuki}}, \citenamefont {{Harada}}, \citenamefont {{Matsuo}}, \citenamefont
  {{Todo}},\ and\ \citenamefont {{Kawashima}}}]{Suzuki2015}%
  \BibitemOpen
  \bibfield  {author} {\bibinfo {author} {\bibfnamefont {Takafumi}\
  \bibnamefont {{Suzuki}}}, \bibinfo {author} {\bibfnamefont {Kenji}\
  \bibnamefont {{Harada}}}, \bibinfo {author} {\bibfnamefont {Haruhiko}\
  \bibnamefont {{Matsuo}}}, \bibinfo {author} {\bibfnamefont {Synge}\
  \bibnamefont {{Todo}}}, \ and\ \bibinfo {author} {\bibfnamefont {Naoki}\
  \bibnamefont {{Kawashima}}},\ }\bibfield  {title} {\enquote {\bibinfo {title}
  {{Thermal phase transition of generalized Heisenberg models for SU (N ) spins
  on square and honeycomb lattices}},}\ }\href {\doibase
  10.1103/PhysRevB.91.094414} {\bibfield  {journal} {\bibinfo  {journal}
  {Physical Review B}\ }\textbf {\bibinfo {volume} {91}},\ \bibinfo {eid}
  {094414} (\bibinfo {year} {2015})},\ \Eprint
  {http://arxiv.org/abs/1505.06273} {arXiv:1505.06273 [cond-mat.stat-mech]}
  \BibitemShut {NoStop}%
\bibitem [{\citenamefont {Cardy}(1996)}]{Cardy1996}%
  \BibitemOpen
  \bibfield  {author} {\bibinfo {author} {\bibfnamefont {John}\ \bibnamefont
  {Cardy}},\ }\href {\doibase 10.1017/CBO9781316036440} {\emph {\bibinfo
  {title} {Scaling and Renormalization in Statistical Physics}}},\ Cambridge
  Lecture Notes in Physics\ (\bibinfo  {publisher} {Cambridge University
  Press},\ \bibinfo {year} {1996})\BibitemShut {NoStop}%
\bibitem [{\citenamefont {Banerjee}\ \emph {et~al.}(2013)\citenamefont
  {Banerjee}, \citenamefont {Jiang}, \citenamefont {Widmer},\ and\
  \citenamefont {Wiese}}]{Banerjee2013}%
  \BibitemOpen
  \bibfield  {author} {\bibinfo {author} {\bibfnamefont {D.}~\bibnamefont
  {Banerjee}}, \bibinfo {author} {\bibfnamefont {F.~J.}\ \bibnamefont {Jiang}},
  \bibinfo {author} {\bibfnamefont {P.}~\bibnamefont {Widmer}}, \ and\ \bibinfo
  {author} {\bibfnamefont {U.~J.}\ \bibnamefont {Wiese}},\ }\bibfield  {title}
  {\enquote {\bibinfo {title} {{The (2 + 1)-d U(1) quantum link model
  masquerading as deconfined criticality}},}\ }\href {\doibase
  10.1088/1742-5468/2013/12/P12010} {\bibfield  {journal} {\bibinfo  {journal}
  {J. Stat. Mech.}\ }\textbf {\bibinfo {volume} {1312}},\ \bibinfo {pages}
  {P12010} (\bibinfo {year} {2013})},\ \Eprint {http://arxiv.org/abs/1303.6858}
  {arXiv:1303.6858 [cond-mat.str-el]} \BibitemShut {NoStop}%
\bibitem [{\citenamefont {{Tschirsich}}\ \emph {et~al.}(2019)\citenamefont
  {{Tschirsich}}, \citenamefont {{Montangero}},\ and\ \citenamefont
  {{Dalmonte}}}]{Tschirsich2019}%
  \BibitemOpen
  \bibfield  {author} {\bibinfo {author} {\bibfnamefont {Ferdinand}\
  \bibnamefont {{Tschirsich}}}, \bibinfo {author} {\bibfnamefont {Simone}\
  \bibnamefont {{Montangero}}}, \ and\ \bibinfo {author} {\bibfnamefont
  {Marcello}\ \bibnamefont {{Dalmonte}}},\ }\bibfield  {title} {\enquote
  {\bibinfo {title} {{Phase diagram and conformal string excitations of square
  ice using gauge invariant matrix product states}},}\ }\href {\doibase
  10.21468/SciPostPhys.6.3.028} {\bibfield  {journal} {\bibinfo  {journal}
  {SciPost Physics}\ }\textbf {\bibinfo {volume} {6}},\ \bibinfo {eid} {028}
  (\bibinfo {year} {2019})},\ \Eprint {http://arxiv.org/abs/1807.00826}
  {arXiv:1807.00826 [cond-mat.stat-mech]} \BibitemShut {NoStop}%
\bibitem [{\citenamefont {Banerjee}(2021)}]{Banerjee2021}%
  \BibitemOpen
  \bibfield  {author} {\bibinfo {author} {\bibfnamefont {Debasish}\
  \bibnamefont {Banerjee}},\ }\bibfield  {title} {\enquote {\bibinfo {title}
  {{Recent progress on cluster and meron algorithms for strongly correlated
  systems}},}\ }\href {\doibase 10.1007/s12648-021-02155-5} {\bibfield
  {journal} {\bibinfo  {journal} {Indian J. Phys.}\ }\textbf {\bibinfo {volume}
  {95}},\ \bibinfo {pages} {1669--1680} (\bibinfo {year} {2021})},\ \Eprint
  {http://arxiv.org/abs/2101.03161} {arXiv:2101.03161 [hep-lat]} \BibitemShut
  {NoStop}%
\bibitem [{\citenamefont {Binder}(1982)}]{Binder1982}%
  \BibitemOpen
  \bibfield  {author} {\bibinfo {author} {\bibfnamefont {K.}~\bibnamefont
  {Binder}},\ }\bibfield  {title} {\enquote {\bibinfo {title} {{Monte Carlo
  calculation of the surface tension for two- and three-dimensional lattice-gas
  models}},}\ }\href {\doibase 10.1103/PhysRevA.25.1699} {\bibfield  {journal}
  {\bibinfo  {journal} {Phys. Rev. A}\ }\textbf {\bibinfo {volume} {25}},\
  \bibinfo {pages} {1699} (\bibinfo {year} {1982})}\BibitemShut {NoStop}%
\bibitem [{\citenamefont {Beach}\ \emph {et~al.}(2005)\citenamefont {Beach},
  \citenamefont {Wang},\ and\ \citenamefont {Sandvik}}]{Beach2005}%
  \BibitemOpen
  \bibfield  {author} {\bibinfo {author} {\bibfnamefont {K.~S.~D.}\
  \bibnamefont {Beach}}, \bibinfo {author} {\bibfnamefont {Ling}\ \bibnamefont
  {Wang}}, \ and\ \bibinfo {author} {\bibfnamefont {Anders~W.}\ \bibnamefont
  {Sandvik}},\ }\bibfield  {title} {\enquote {\bibinfo {title} {Data collapse
  in the critical region using finite-size scaling with subleading
  corrections},}\ }\href {\doibase 10.48550/ARXIV.COND-MAT/0505194} {\
  (\bibinfo {year} {2005}),\ 10.48550/ARXIV.COND-MAT/0505194}\BibitemShut
  {NoStop}%
\bibitem [{\citenamefont {Wang}\ \emph {et~al.}(2006)\citenamefont {Wang},
  \citenamefont {Beach},\ and\ \citenamefont {Sandvik}}]{Wang2006}%
  \BibitemOpen
  \bibfield  {author} {\bibinfo {author} {\bibfnamefont {Ling}\ \bibnamefont
  {Wang}}, \bibinfo {author} {\bibfnamefont {K.~S.~D.}\ \bibnamefont {Beach}},
  \ and\ \bibinfo {author} {\bibfnamefont {Anders~W.}\ \bibnamefont
  {Sandvik}},\ }\bibfield  {title} {\enquote {\bibinfo {title} {High-precision
  finite-size scaling analysis of the quantum-critical point of $s=1/2$
  heisenberg antiferromagnetic bilayers},}\ }\href {\doibase
  10.1103/PhysRevB.73.014431} {\bibfield  {journal} {\bibinfo  {journal} {Phys.
  Rev. B}\ }\textbf {\bibinfo {volume} {73}},\ \bibinfo {pages} {014431}
  (\bibinfo {year} {2006})}\BibitemShut {NoStop}%
\bibitem [{\citenamefont {Hasenbusch}(2010)}]{Hasenbusch2010}%
  \BibitemOpen
  \bibfield  {author} {\bibinfo {author} {\bibfnamefont {Martin}\ \bibnamefont
  {Hasenbusch}},\ }\bibfield  {title} {\enquote {\bibinfo {title} {{Finite size
  scaling study of lattice models in the three-dimensional Ising universality
  class}},}\ }\href {\doibase 10.1103/PhysRevB.82.174433} {\bibfield  {journal}
  {\bibinfo  {journal} {Phys. Rev. B}\ }\textbf {\bibinfo {volume} {82}},\
  \bibinfo {pages} {174433} (\bibinfo {year} {2010})},\ \Eprint
  {http://arxiv.org/abs/1004.4486} {arXiv:1004.4486 [cond-mat.stat-mech]}
  \BibitemShut {NoStop}%
\bibitem [{\citenamefont {Jos\'e}\ \emph {et~al.}(1977)\citenamefont {Jos\'e},
  \citenamefont {Kadanoff}, \citenamefont {Kirkpatrick},\ and\ \citenamefont
  {Nelson}}]{Jose1977}%
  \BibitemOpen
  \bibfield  {author} {\bibinfo {author} {\bibfnamefont {Jorge~V.}\
  \bibnamefont {Jos\'e}}, \bibinfo {author} {\bibfnamefont {Leo~P.}\
  \bibnamefont {Kadanoff}}, \bibinfo {author} {\bibfnamefont {Scott}\
  \bibnamefont {Kirkpatrick}}, \ and\ \bibinfo {author} {\bibfnamefont
  {David~R.}\ \bibnamefont {Nelson}},\ }\bibfield  {title} {\enquote {\bibinfo
  {title} {Renormalization, vortices, and symmetry-breaking perturbations in
  the two-dimensional planar model},}\ }\href {\doibase
  10.1103/PhysRevB.16.1217} {\bibfield  {journal} {\bibinfo  {journal} {Phys.
  Rev. B}\ }\textbf {\bibinfo {volume} {16}},\ \bibinfo {pages} {1217--1241}
  (\bibinfo {year} {1977})}\BibitemShut {NoStop}%
\bibitem [{\citenamefont {Delfino}\ and\ \citenamefont
  {Lamsen}(2019)}]{Delfino2019}%
  \BibitemOpen
  \bibfield  {author} {\bibinfo {author} {\bibfnamefont {Gesualdo}\
  \bibnamefont {Delfino}}\ and\ \bibinfo {author} {\bibfnamefont {Noel}\
  \bibnamefont {Lamsen}},\ }\bibfield  {title} {\enquote {\bibinfo {title}
  {Critical points of coupled vector-ising systems. exact results},}\ }\href
  {\doibase 10.1088/1751-8121/ab3055} {\bibfield  {journal} {\bibinfo
  {journal} {Journal of Physics A: Mathematical and Theoretical}\ }\textbf
  {\bibinfo {volume} {52}},\ \bibinfo {pages} {35LT02} (\bibinfo {year}
  {2019})}\BibitemShut {NoStop}%
\bibitem [{\citenamefont {Marcos}\ \emph {et~al.}(2014)\citenamefont {Marcos},
  \citenamefont {Widmer}, \citenamefont {Rico}, \citenamefont {Hafezi},
  \citenamefont {Rabl}, \citenamefont {Wiese},\ and\ \citenamefont
  {Zoller}}]{Marcos2014}%
  \BibitemOpen
  \bibfield  {author} {\bibinfo {author} {\bibfnamefont {D.}~\bibnamefont
  {Marcos}}, \bibinfo {author} {\bibfnamefont {P.}~\bibnamefont {Widmer}},
  \bibinfo {author} {\bibfnamefont {E.}~\bibnamefont {Rico}}, \bibinfo {author}
  {\bibfnamefont {M.}~\bibnamefont {Hafezi}}, \bibinfo {author} {\bibfnamefont
  {P.}~\bibnamefont {Rabl}}, \bibinfo {author} {\bibfnamefont {U.~J.}\
  \bibnamefont {Wiese}}, \ and\ \bibinfo {author} {\bibfnamefont
  {P.}~\bibnamefont {Zoller}},\ }\bibfield  {title} {\enquote {\bibinfo {title}
  {{Two-dimensional Lattice Gauge Theories with Superconducting Quantum
  Circuits}},}\ }\href {\doibase 10.1016/j.aop.2014.09.011} {\bibfield
  {journal} {\bibinfo  {journal} {Annals Phys.}\ }\textbf {\bibinfo {volume}
  {351}},\ \bibinfo {pages} {634--654} (\bibinfo {year} {2014})},\ \Eprint
  {http://arxiv.org/abs/1407.6066} {arXiv:1407.6066 [quant-ph]} \BibitemShut
  {NoStop}%
\bibitem [{\citenamefont {Celi}\ \emph {et~al.}(2020)\citenamefont {Celi},
  \citenamefont {Vermersch}, \citenamefont {Viyuela}, \citenamefont {Pichler},
  \citenamefont {Lukin},\ and\ \citenamefont {Zoller}}]{Celi2019}%
  \BibitemOpen
  \bibfield  {author} {\bibinfo {author} {\bibfnamefont {Alessio}\ \bibnamefont
  {Celi}}, \bibinfo {author} {\bibfnamefont {Beno\^\i{}t}\ \bibnamefont
  {Vermersch}}, \bibinfo {author} {\bibfnamefont {Oscar}\ \bibnamefont
  {Viyuela}}, \bibinfo {author} {\bibfnamefont {Hannes}\ \bibnamefont
  {Pichler}}, \bibinfo {author} {\bibfnamefont {Mikhail~D.}\ \bibnamefont
  {Lukin}}, \ and\ \bibinfo {author} {\bibfnamefont {Peter}\ \bibnamefont
  {Zoller}},\ }\bibfield  {title} {\enquote {\bibinfo {title} {{Emerging
  Two-Dimensional Gauge Theories in Rydberg Configurable Arrays}},}\ }\href
  {\doibase 10.1103/PhysRevX.10.021057} {\bibfield  {journal} {\bibinfo
  {journal} {Phys. Rev. X}\ }\textbf {\bibinfo {volume} {10}},\ \bibinfo
  {pages} {021057} (\bibinfo {year} {2020})},\ \Eprint
  {http://arxiv.org/abs/1907.03311} {arXiv:1907.03311 [quant-ph]} \BibitemShut
  {NoStop}%
\end{thebibliography}%
\newpage
\appendix 

\section{Supplementary Material}
\subsection{Symmetries with and without static $\pm 2$ charges} 
\label{subsect1}
 In this section, we discuss the various symmetries of the model with and without the $\pm 2$ charges. 
 Let us first discuss the case without the charges, where one has the usual lattice symmetries of 
 translations by one lattice spacing, the various reflection and rotation symmetries which form the 
 point group symmetries. The charge conjugation is a global internal $Z_2$ symmetry, which transforms
 as: $U \rightarrow U^\dagger, U^\dagger \rightarrow U, E \rightarrow -E$. Additionally, there is the 
 global symmetry associated with the large gauge transformations, generated by the operator 
 $W_i = \frac{1}{L_i} \sum_{x} E_{x,i}$ and classifies the physical states in winding sectors taking 
 values in ${\mathbb{Z}}$ for even $L$. Finally, the gauge symmetry, corresponding to the choice of the Gauss Law 
 allows only the six allowed states with zero charge at a vertex.
 
\begin{figure}
     \centering
     \includegraphics[scale=0.9]{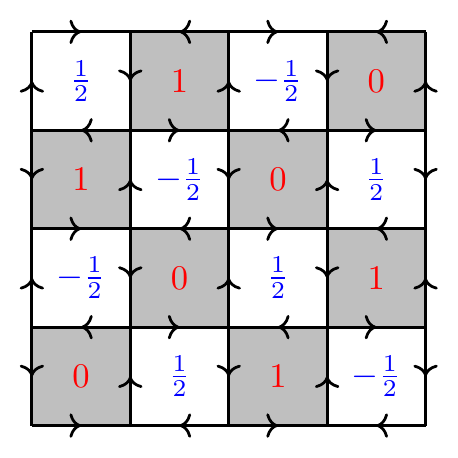}
     \includegraphics[scale=0.9]{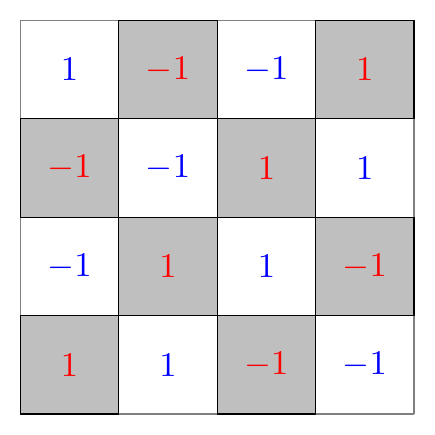}
     \caption{(Left): Reference configuration for a $L=4$ lattice. Such a configuration has the maximum
     number of flippable plaquettes, which is signalled by large values of both $M_{A},M_{B}$. Both the
     flux (on the links) and the height variables (in the dual lattices) are shown. (Right) The phase
     factors $\eta^{X}_{\tilde{x}}$ for the A and B sublattices which need to be multiplied with the
     height variables $h^X$ to obtain the magnetization.}
     \label{fig:refCL4}
 \end{figure}

  We note that the QLM with spin $S = \frac{1}{2}$ has a different ground state phase diagram from the
 Wilsonian LGT. The latter uses quantum rotors as degrees of freedom, and thus the electric flux are 
 quantized in integer units, whereas in our case the electric flux is always $\pm \frac{1}{2}$. The 
 QLM ground state diagram has three distinct phases: two different confined phases (C1 and C2), and 
 the staggered phase. For large negative $\lambda$, where phase C1 is stabilized, the ground state 
 has the maximal number of flippable plaquettes (see \Cref{fig:refCL4} (left)). This phase spontaneously 
 breaks lattice translation symmetry by one lattice spacing, as well as charge conjugation symmetry. 
 As $\lambda$ is reduced, the $J$ term governing the flips dominate, and the system goes into a new 
 phase (C2) where one of the two sublattices (for example, the shaded one) is in a coherent 
 superposition of clockwise and anticlockwise flippable plaquettes, often called a resonating valence 
 bond (RVB) solid. C2 has an unbroken charge conjugation symmetry, but a still broken lattice 
 translation symmetry. In terms of magnetization operators (introduced in the Main text), in the 
 phase C1 both sublattices are ordered, while C2 has ordering of one of the two sublattices.
 
  When the $\pm 2$ charges are included in the thermal ensemble (by appropriately modifying the Gauss 
 Law), several of the previous global symmetries do not remain good quantum numbers any more. Since the 
 distribution of charges in the volume is governed thermally, lattice translation, rotation and reflections 
 are not good symmetries. Charge conjugation is explicitly broken in the presence of charges. The total 
 number of allowed states at a vertex is now 8, out of the 16 possible states in total. The other 8 
 states correspond to the presence of a $\pm 1$ charge, and are realized in the quantum dimer model, but 
 not in our model. With 8 vertex states allowed by the Gauss Law, this is an effective $Z_2$
 gauge constraint on the Hamiltonian in \Cref{eq:ham}. It can also be viewed as a quantum eight-vertex
 model. A more mathematical argument is provided below to make the effective $Z_2$ nature of the Gauss Law 
 explicit. The winding numbers now take values in $\mathbb{Z}/2$ for even $L$. 
 
  Since the Gauss law for the pure gauge theory annihilates the physical states $\ket{\psi}$, it follows
 that the action of an arbitrary gauge transformation on the physical states keeps them unchanged:
\begin{equation}
\begin{split}
  V\ket{\psi} & = \prod_x \exp (i\theta_x G_x) \ket{\psi}  \\
     & = \prod_x \left[ 1 + i \theta_x G_x - \frac{\theta_x^2 G_x^2}{2} + \cdots \right] \ket{\psi}
       = \ket{\psi}.
\end{split}
\end{equation}
 The $U(1)$ nature of the gauge symmetry is because $\theta_x \in (0, 2 \pi]$ is an angle. 
When we allow $Q_x = 0, \pm 2$ in our theory, then we have two classes of states, 
$\ket{\Psi} = \left\{ \ket{\psi}, \ket{\phi} \right\} $, which satisfy, $G_x \ket{\psi} = 0$
and $G_x \ket{\phi} = \pm 2 \ket{\phi}$. Under the same gauge transformation as before, we now have
 \begin{equation}
 \begin{split}
		V\ket{\Psi} & = \prod_x \exp (i\theta_x G_x) \ket{\Psi} \\
 		         	& = \prod_x \left[ 1 + i \theta_x G_x -\frac{\theta_x^2 G_x^2}{2}+  \cdots \right] \ket{\Psi}
 \end{split}
 \end{equation}
 For the states $\ket{\psi}$ it is clearly satisfied, but for the states $\ket{\phi}$, the states are only 
 unchanged when $\theta_x = 0, \pi$. Thus only a $Z_2$ subgroup of the original $U(1)$ survives when we 
 demand gauge invariance for the states having $\pm 2$ charges in addition to the zero charged ones.
 (Note that the total charge is still zero, we are only referring to the local charges). 
 
 \begin{figure}
     \centering
     \includegraphics[scale=1.0]{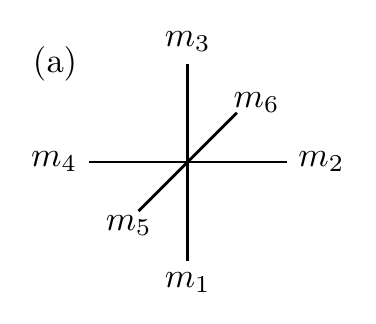} \\
     \includegraphics[scale=0.8]{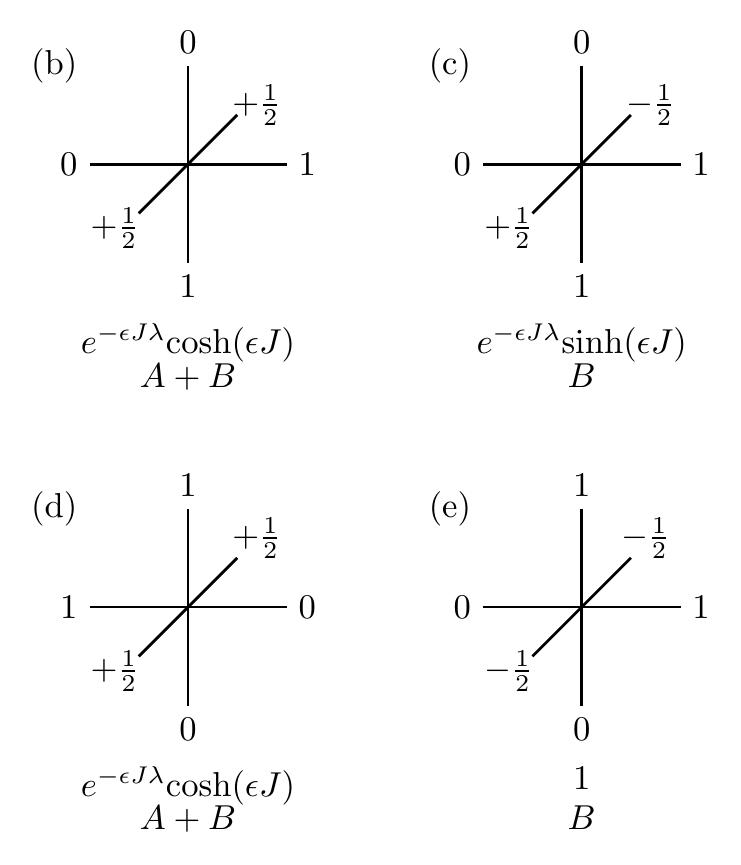}
     \caption{The layout of the height variables and the cluster rules.
     (a) $m_1, m_2, m_3, m_4$ lie on a timeslice $t$, $m_5$ is on timeslice $t-1$, and $m_6$ on timeslice
     $t+1$. (b) and (c) Out-of-plane breakups: when there is a reference configuration, one can connect
     $m_5$ and $m_6$ with a certain probability only if they are the same ($A$-breakup). The $B$-breakup
     corresponds to the case when $m_5 \neq m_6$ and they cannot be connected.
     (d) and (e) In-plane breakups: If $m_5 \neq m_6$, $m_1,m_2,m_3,m_4$ are in a reference configuration
     and must be connected, while for $m_5 = m_6$ the connection of $m_1,m_2,m_3,m_4$ is done with only
     a certain probablity.
     }
     \label{fig:QLMC}
 \end{figure}

\subsection{Dualization, quantum Monte Carlo algorithm, and order parameters}
  Construction of efficient quantum Monte Carlo algorithms for (lattice) gauge theories is a particularly
 difficult challenge due to the constraints that need to be satisfied. Stochastic local updates typically 
 get rejected when they cannot satisfy the constraints, while global updates satisfying detailed balance
 are non-trivial to construct. Our model, being a lattice gauge theory is no exception. However, recently
 it has become possible to exploit dualization techniques in order to rewrite the problem in terms of 
 different variables which partially solve the gauge constraints. The $U(1)$ theory in 
 $(2+1)-$d can be dualized into a $(2+1)-$d quantum height model, for which efficient cluster algorithms
 can be constructed. In particular, rewriting the gauge theory in terms of the height variables 
 completely eliminates the odd charges $Q_x = \pm 1$ from the theory written in terms of the height
 variables. The resulting theory only allows $Q_x = 0, \pm 2$, which explains the
 algorithmic ease of incorporating the $\pm 2$ charges. However, as was done in \cite{Banerjee2013}, it is easy
 to impose a further constraint in the cluster algorithm to project out the $\pm 2$ charges while studying
 the pure gauge theory. We provide an outline of the procedure below, for further details please see 
 \cite{Banerjee2021}.
 
 \begin{figure}
     \includegraphics[scale=0.8]{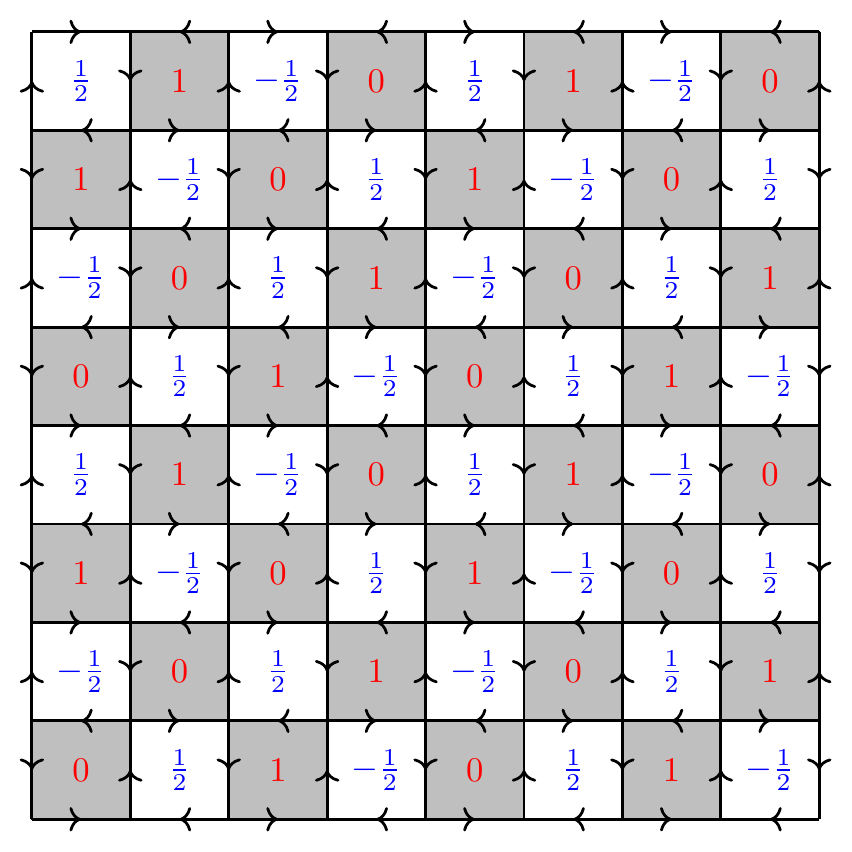}
     \includegraphics[scale=0.8]{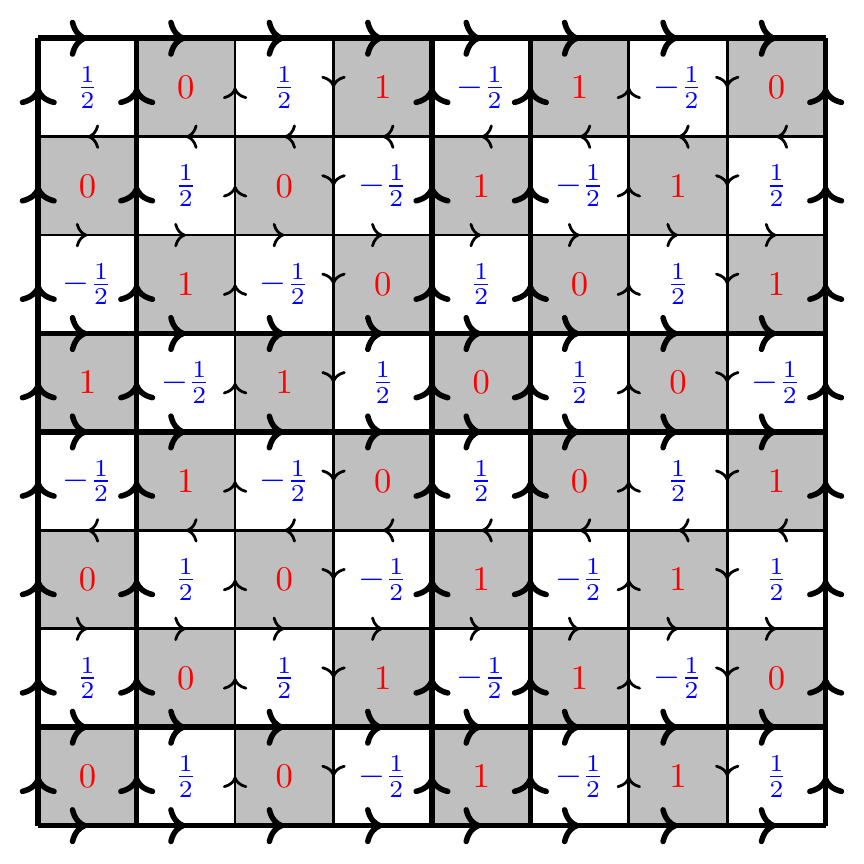}
     \caption{(top) A sample configuration on the $L=8$ lattice with the largest magnetization for both 
     $M_A$ and $M_B$, which corresponds to all flippable plaquettes. Both the flux and the height 
     configurations are shown. For this configuration, $M_A = -16$ and $M_B = 16$. 
     (bottom) A different configuration for the $L=8$ lattice with winding numbers $W_x = W_y= 2$, showing 
     both the height and the flux configurations, for which $M_A = M_B = 0$. The links carrying the
     winding strings, both in the horizontal and vertical directions, are shown in bold.}
     \label{fig:refCL8}
 \end{figure}
 
  \Cref{eq:ham} and \Cref{eq:pf} display the Hamiltonian and the partition function of our model, respectively.
 To proceed further, we construct the transfer matrix $\mathbb{T}=e^{-\epsilon H}$, by discretizing the temporal direction 
 into $L_T$ time slices ($\beta = \epsilon L_T$). Moreover, noting that each link participates in two plaquette
 interactions, it is natural to divide the lattice into even ($A$) and odd ($B$) sublattices, which is the well-known
 Trotter decomposition. All the plaquettes in sublattice A (or sublattice B) can be considered to be interacting 
 simultaneously. \Cref{fig:refCL4} shows the division into two sublattices A and B (shaded and unshaded) respectively. 
 The Hamiltonian and partition function can now be written as,
 \begin{equation}
   H = H_A + H_B \;\;\; \text{and}\;\;\; Z=\text{Tr}\left[ \mathbb{T}_A^{L_T}\mathbb{T}_B^{L_T} P_{G}\right]  
 \end{equation}
 where $\mathbb{T}_{A(B)}=\text{exp}[-\epsilon H_{A(B)}]$ and we have neglected $O(\epsilon ^2)$ terms. $P_G$ is 
 the projection operator enforcing the Gauss law: 
 $P_G = \prod_x \{ 6\delta(G_x) + \delta(G_x - 2) + \delta(G_x + 2)\} \frac{1}{8}$ when we need to include
 the charge $\pm 2$, or $P_G = \prod_x \delta (G_x)$, for the case without the charges. We can read off the 
 weights for the various configurations from the single plaquette transfer matrix operator:
 \begin{equation}
 \begin{split}
 \mathbb{T}_\square & = 1 + (U_\square + U_\square^\dagger) {\rm e}^{-\epsilon\lambda J} \text{sinh}(\epsilon J) \\
  & + (U_\square + U_\square^\dagger)^2 [{\rm e}^{-\epsilon \lambda J}\text{cosh}(\epsilon J) - 1 ]    
 \end{split}
 \end{equation}
  The non flippable plaquettes (fourteen of them) have unit weights. The two flippable plaquettes give diagonal 
  contribution of  $e^{-\epsilon\lambda J} \text{cosh} (\epsilon J)$. A configuration where a plaquette is flipped 
  has a weight $e^{-\epsilon\lambda J}\text{sinh}(\epsilon J)$ (the off-diagonal elements).
  
  On dualization, the model is reformulated in terms of quantum height variables, which live in the 
 centres of the plaquettes (the dual lattice sites). Thus, if a site is labelled as $x = (x_1,x_2)$, 
 the dual sites (where the dual heights are located) are at $\tilde{x}=(x_1 + \frac{1}{2}, x_2 + \frac{1}{2})$. 
 Moreover, the height variables on different sublattices are distinct: $h_{\tilde{x}}^{\rm A}=0,1$ for even 
 (A) sublattice and $h_{\tilde{x}}^{\rm B}=\pm \frac{1}{2}$ for odd (B) sublattice. The divergence of the 
 height variables are the electric flux variables, 
 \begin{equation}
  E_{x,\h{i}}= \left[ h_{\tilde{x}}^X-h_{\tilde{x}+\tilde{i}-\tilde{1}-\tilde{2}}^{X^\prime} \right]
  ~~{\rm mod2} = \pm\frac{1}{2}; ~~~ {\rm X,X}^\prime \in \{{\rm A},{\rm B}\}    
 \end{equation}
 An example of the flux to height mapping is shown in \Cref{fig:refCL4} (left). We note that the above 
 relation remains unchanged when both the height variables are flipped, and thus there are two distinct 
 height configurations which map to a single flux configuration. Moreover, it is easy to check by 
 constructing the height variables in the presence of charges $Q_x = \pm 1$ that the height assignment 
 does not work out, while $Q_x = \pm 2$ does not present a problem. In other words, the dualization using 
 the height variables projects out the $Q_x = \pm 1$.
 
 The cluster algorithm works by building clusters in a single sublattice at a time. Bonds between the 
 variables in a sublattice (say A) are put depending on the value of the height variables in the other 
 sublattice (therefore B), which we now discuss. In \Cref{fig:QLMC} (a), the general layout of height 
 variables across time slices is shown: $m_1,\:m_2\:,m_3$ and $m_4$ lie in time slice $t$,  $m_5$ lies 
 in time slice $t-1$ and $m_6$ lies in time slice $t+1$. We shall consider two different breakups, 
 the out-of-plane breakups (whether to bond $m_5$ and $m_6$) and in-plane breakups (whether to bond 
 $m_1,\:m_2,\:m_3$ and $m_4$). If we can bond heights, we call it a $A$-breakup; otherwise it is a 
 $B$-breakup.

 \begin{itemize}
     \item \textbf{Out-of-plane breakups}: 
     \begin{itemize}
         \item If ($m_1,m_2,m_3,m_4$) are not in reference configuration, $m_5$ must be equal to $m_6$, and 
         we must bond ($m_5,m_6$) to forbid disallowed configurations. 
         \item If ($m_1,m_2,m_3,m_4$) are in reference configuration, either $m_5\neq m_6$ (weight is 
         $e^{-\epsilon\lambda J}\sinh(\epsilon J)$) or $m_5=m_6$ (weight is $e^{-\epsilon\lambda J} 
         \cosh(\epsilon J)$). If $m_5\neq m_6$, have case $B$ (see \Cref{fig:QLMC} (c)). If $m_5=m_6$, 
         both $A$ and $B$ can happen (see \Cref{fig:QLMC}(b)). To satisfy detailed balance, we have, 
         $P_B=B/(A+B)=\tanh(\epsilon J)$ and $P_A=1-P_B=\frac{e^{-\epsilon J}}{\cosh(\epsilon J)}$.
     \end{itemize}
     \item \textbf{In-plane breakups}: 
      \begin{itemize}
       \item If $m_5\neq m_6$, ($m_1,m_2,m_3,m_4$) are in a reference configuration and we must bind 
       them together to avoid disallowed configuration. 
       \item If $m_5=m_6$, either ($m_1,m_2,m_3,m_4$) are in reference configuration (with weight 
       $e^{-\epsilon\lambda J}\cosh(\epsilon J)$) or they are not (weight is $1$). If ($m_1,m_2,m_3,m_4$) 
       form a reference configuration, both $A$ and $B$ can be applied, else, we are in case $B$ 
       (see \Cref{fig:QLMC}). Solving the detailed balance equation we get, 
       $P_B=B/(A+B)=\frac{e^{\epsilon\lambda J}}{\cosh(\epsilon J)}$ and 
       $P_A=1- \frac{e^{\epsilon\lambda J}}{\cosh(\epsilon J)}$.
      \end{itemize}
 \end{itemize}
 The cluster building process proceeds as is usual in a spin model. It is possible to consider the
 single cluster Wolff algorithm, or the multi-cluster Swendsen-Wang variant. For our case, we have
 implemented both the algorithm and checked that the answers match.
 
  Another important point is the construction of order parameters, $M_{A}$ and $M_B$, which are constructed
  as $M_{X} = \frac{1}{L_T}\sum_{\tilde{x}} \eta^X_{\tilde{x}} h^X_{\tilde{x}}$, where $X=A,B$ and $\tilde{x}$ 
  denote the dual sites (centres of the plaquettes of the original space-time lattice). The phases $\eta^X$,
  which need to be multiplied with the height variables $h^X$ to get the magnetization, are displayed in 
  \Cref{fig:refCL4} (right) for $L=4$, but for any other lattice (with a multiple of 4), they can be easily tiled.
  We note that this order parameter is sensitive to the different confined phases: C1 and C2. While C1 has 
  ordering on both sublattices (and hence have maximum values of both $M_A$ and $M_B$), C2 has order on only
  one sublattice (so either $M_A$ or $M_B$ is maximum and the other is zero). An example of the C1 phase
  is shown for the $L=8$ in \Cref{fig:refCL8} (top) with both the height and the flux configurations,
  for which one has $M_A = -16$ and $M_B = 16$. Interestingly, these order parameters also signal deconfinement 
  by \emph{both} going to zero simultaneously, when the lattice volume has many (extensive) winding strings. An example of 
  a state with winding strings is shown in \Cref{fig:refCL8} (bottom), for which $M_A = M_B = 0$.

\subsection{Methods: Exact diagonalization and cluster QMC}
  For our model in two spatial dimensions ($L_X,L_Y$) and spin-$\frac{1}{2}$ degrees of freedom per link, 
  there are $2^{2 L_X L_Y}$ possible configurations, while working in the electric flux basis. Imposing 
  the Gauss law constraint (which is easy in the flux basis) greatly reduces the number of allowed basis 
  states in the Hilbert space. For example, a (4,4) lattice with only zero-charges has 2970 basis states,
  while the same lattice allowing for additional $\pm 2$-charges has 131072 basis states. 
  
  A pragmatic way to proceed is to divide the Hilbert space into various winding number sectors ($W_x,W_y$). 
  Since these sectors do not mix up among themselves, we can independently work in each individual sector, 
  thus effectively reducing the Hilbert space dimension. In the presence of only $Q=0$ the individual 
  winding numbers along $x$ or $y$ direction are all the same irrespective of where they are summed. 
  Now, if we allow $Q = \pm 2$ along with $Q=0$, the individual winding numbers along any direction can, 
  in principle, be different; so we should label these sectors as $(\{W_{x,1},W_{x,2},...,W_{x,L_Y}\},
  \{W_{y,1},W_{y,2},...,W_{y,L_X}\})$. Interestingly, all the winding numbers along a direction,
  are either all even or all odd. Thus, there are four sectors: (even, even), (even, odd), (odd, even) 
  and (odd, odd). For a (4,4) lattice, the winding numbers can take values 0, $\pm1$ and $\pm2$. So 
  there are $3^4 \times 3^4$ sub-sectors within (even, even) sector, $3^4\times 2^4$ sub-sectors 
  within (even, odd) or (odd, even) sector and $2^4\times 2^4$ sub-sectors within (odd, odd) sector. 
  Thus, the Hilbert space in a given winding sector is further divided into many small sub-sectors which 
  do not mix among themselves, facilitating ED significantly. For a (4,4) lattice, in the presence of 
  only zero charges we need to diagonalize the largest sector (0,0) which has 990 basis states, whereas, 
  in the addition of $\pm2$ charges we need to diagonalize the largest sector ($\{0,0,0,0\},\{0,0,0,0\}$) 
  with 3464 basis states.

  Our QMC cluster algorithm can simulate different winding number sectors, but the height representation 
  constrains the winding sectors that can be simulated at a time. For example, using periodic boundary
  conditions, only even winding sectors can be simulated, while for odd winding sectors one would need
  anti-periodic boundary conditions. While this is irrelevant for infinite volume in the spatial directions,
  it is important in the shorter temporal directions to have periodic boundary conditions on the height
  variables. Periodic boundary conditions on the flux variables translate to both \emph{periodic} and
  \emph{anti-periodic} boundary conditions on the height variables in time. However, the latter gives
  rise to an interface, which would become energetically unfavourable when one approaches the thermodynamic
  limit. Thus, it is completely natural to use the QMC on the height variables and restrict oneself to the
  even winding sectors. 
  
\begin{figure}
     \centering
     \includegraphics[width=\hsize]{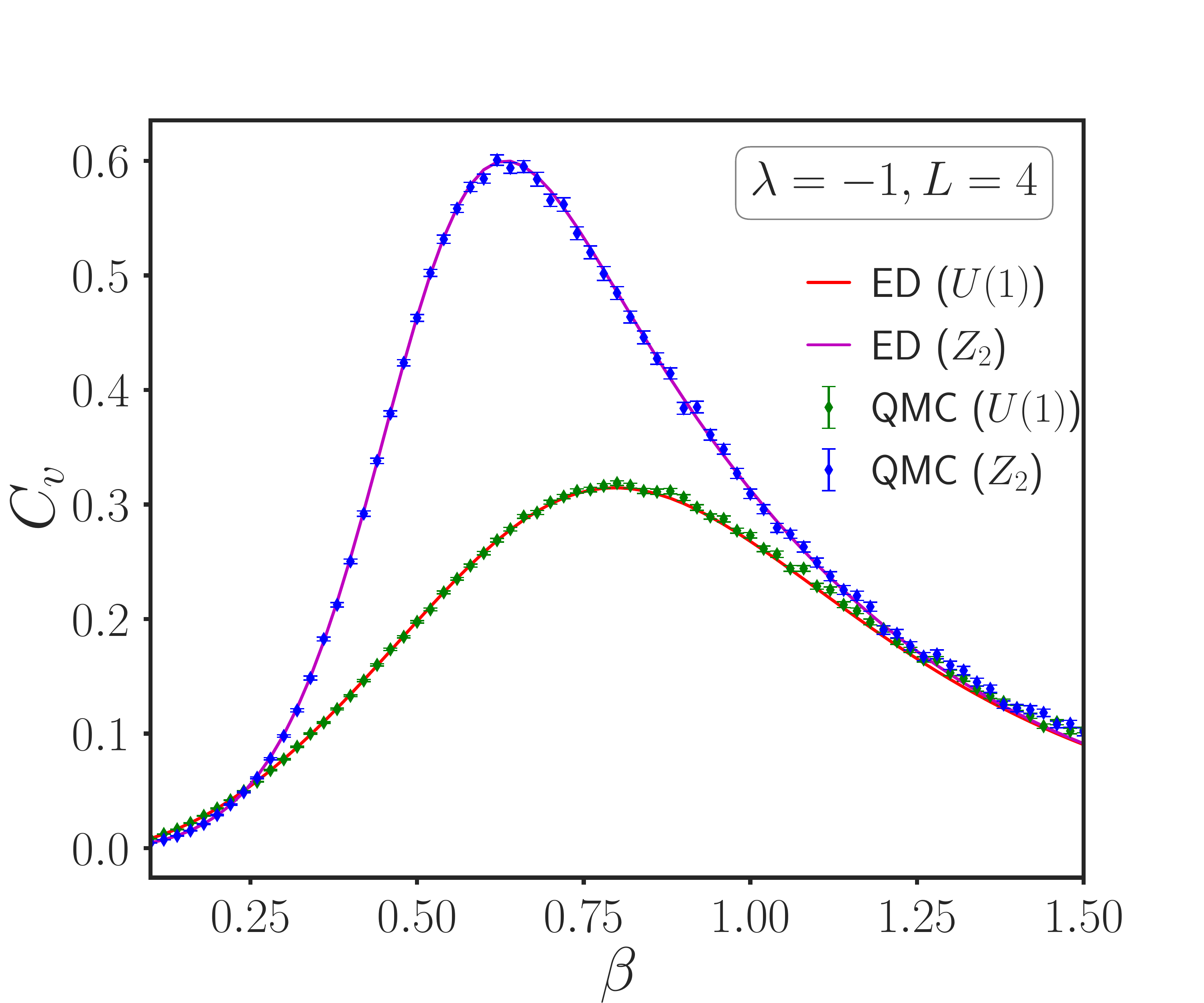}
     \caption{Comparison of ED and QMC.}
     \label{fig:Z2QLMC}
 \end{figure}

  In \Cref{fig:Z2QLMC} we have plotted specific heat, $C_v=\beta^2(\braket{E^2} - \braket{E}^2)$ 
  versus $\beta$, obtained using QMC and ED for both absence and presence of $Q = \pm 2$. Here 
  $\braket{E} = -\frac{\partial \textrm{ln}Z}{\partial \beta}=-\frac{1}{L_T}\frac{\partial \rm{ln}Z}{\partial \epsilon}$ 
  is average energy and $\braket{E^2} = \frac{1}{Z}\frac{\partial^2 Z}{\partial\beta^2}=\frac{1}{Z L_T^2}\frac{\partial^2
  Z}{\partial\epsilon^2}$ is the average of energy squared, where the partition function is 
  $Z=\sum_{\rm{config.}}\prod_\square W_\square(\epsilon J,\epsilon \lambda)$. 
  $W_\square$ is the weight associated with a certain 6-height-variables interaction. $W_\square$ can take 
  values $W_\square^1 = e^{-\epsilon J\lambda}\sinh(\epsilon J)$ for a flipped plaquette, 
  $W_\square^2 = e^{-\epsilon J\lambda}\cosh(\epsilon J)$ for an unchanged flippable plaquette 
  and $W_\square^3=1$ otherwise. After simplification, 
  $\braket{E} = -\frac{1}{L_T}\braket{ \sum_\square\frac{\partial \rm{ln}W_\square }{\partial \epsilon} }$ 
  and $\braket{ E^2}$ reads as 
  $\frac{1}{L_T^2} \left( \braket{\sum_\square \frac{\partial^2 \rm{ln}W_\square}{\partial \epsilon^2}}  
  + \braket{\sum_\square (\frac{\partial \rm{ln}W_\square}{\partial \epsilon})^2} \right)$.
  To compute $\braket{E}$ and $\braket{E^2}$ using QMC, we need to go through all 6-height-variables 
  interactions and check the corresponding weights associated with them and add the appropriate factors 
  in accordance with the simplified forms of $\braket{E}$ and $\braket{E^2}$.

  We note the nice agreement of results between QMC and ED in \Cref{fig:Z2QLMC}, which indicates the 
  efficiency of the cluster algorithm we have used.
 
 \begin{figure}
    \includegraphics[width=\hsize]{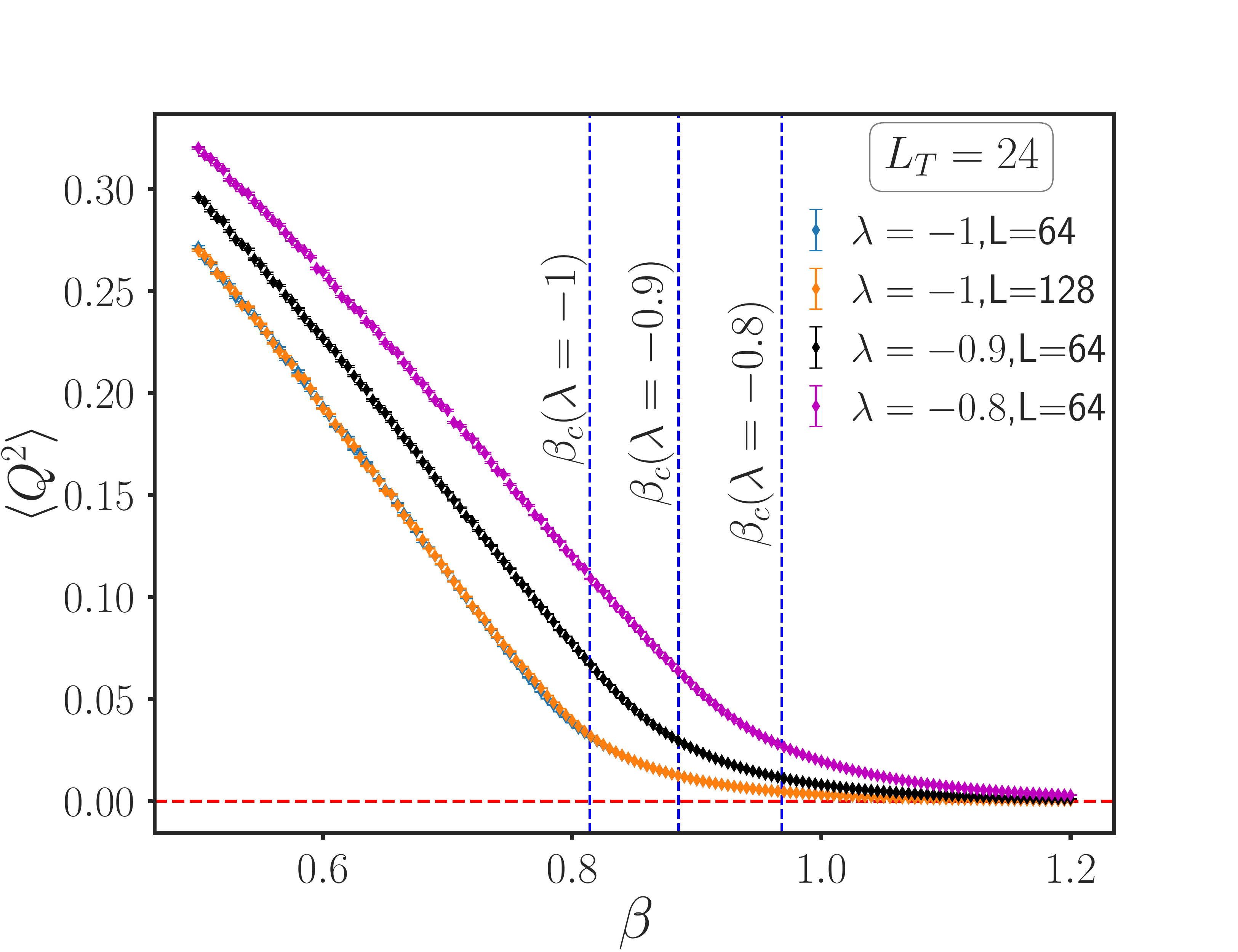}
    \caption{The $\braket{Q^2}$ operator for different values of $\lambda$ for a
    range of $\beta$ on our finest lattice, $L_T = 24$. The vertical dotted line
    shows the location of critical temperature for the different $\lambda$ values.}
    \label{fig:Qsqr}
\end{figure}

 \subsection{Variation of local charge density with temperature}
  While the $U(1)$ QLM has $\braket{Q^2}=0$ (normalized with the spatial volume) at all $T$ and $\lambda$ 
  by definition, this is not the case for the LGT defined by the partition function in \Cref{eq:pf}. For 
  example, it is easy to see that $\braket{Q^2}\rightarrow 1$ as $T \rightarrow \infty$ from \Cref{eq:pf}. 
  On the other hand, the mass gap of $Q=\pm 2$ charges scale as $O(|\lambda|)$ from \Cref{eq:ham} for 
  negative values of $\lambda$ which implies that $\braket{Q^2} \sim \exp(-a |\lambda|/T)$, where $a$ 
  is an $O(1)$ number, as $T \rightarrow 0$.

  In Fig.~\ref{fig:Qsqr}, we show the variation of $\braket{Q^2}$ as a function of $\beta$ from the QMC 
  data for the $U(1)$ QLM with $Q=\pm 2$ (\Cref{eq:pf}) for $\lambda=-1.0, -0.9,-0.8$ respectively using 
  $L_T=24$. At low $T$ and also in the neighborhood of $T_c=1/\beta_c$, $\braket{Q^2}$ follows an activated 
  behavior of $\exp(-a |\lambda|/T)$ for all the three values of $\lambda$. Interestingly, even though 
  $\braket{Q^2} \approx 0.03$ in the vicinity of $\beta_c$ and thus the $Q=\pm 2$ charges can be considered 
  to be dilute at the deconfinement transition, this thermal population is sufficient to completely change 
  the critical behavior compared to the case where $\braket{Q^2}$ is strictly zero. We thus conjecture that 
  these charged degrees of freedom ($Q=\pm 2$) generate a marginal operator and hence, weak universality in this LGT. 
\end{document}